\documentclass[fleqn,11pt,a4paper]{article}
\usepackage[dvips]{graphicx}

\newcommand{\lsim}{\rlap{\raise 2pt \hbox{$<$}}{\lower 2pt \hbox{$\sim$}}}
\newcommand{\gsim}{\rlap{\raise 2pt \hbox{$>$}}{\lower 2pt \hbox{$\sim$}}}
\newcommand{\earth}{{\oplus}}
\newcommand{\etal}{{\it et al.}}
\newcommand{\ie}{{\it i.e.\ }}
\newcommand{\eg}{{\it e.g.\ }}

\newcommand{\ccbar}{$c\bar{c}$}                                                 
\renewcommand{\d}{\mbox{\rm d}}

\begin{document}


\thispagestyle{empty}
\noindent
TSL/ISV-95-0120  \hfill ISSN 0284-2769 \\
PAR-LPTHE-95-29 \\
May 1995, revised February 1996  \\
\begin{center}
\begin{LARGE}
\begin{bf}
Charm Production and High Energy\\ Atmospheric Muon and Neutrino Fluxes\\
\end{bf}
\end{LARGE}
\vspace{0.5cm}
\begin{Large}
M.~Thunman\footnote{thunman@tsl.uu.se}, G.~Ingelman\footnote{
ingelman@tsl.uu.se, also at DESY, Hamburg.} \\
\end{Large}
\vspace{0.2cm}
{\it Department of Radiation Sciences,  Uppsala University,
 Box 535, S-751 21 Uppsala, Sweden }\\
\vspace{0.2cm}

\begin{Large}
P.~Gondolo\footnote{Present address: Oxford University, Dept. of Physics, 1
Keble Road, OX1 3NP, UK,\\ \indent $\ \,$ p.gondolo1@physics.oxford.ac.uk} \\
\end{Large}
\vspace{0.2cm}
{\it Physique Th\'eorique et Hautes Energies,
 Universit\'e de Paris 7, Boite 7109, 2 place Jussieu, \\
\hspace*{2mm} F-75251 Paris, France}\\

\end{center}

%
\begin{quotation}
\noindent
{\bf Abstract:}
Production of muons and neutrinos in cosmic ray interactions with the 
atmosphere has been investigated with Monte Carlo models  for hadronic
interactions. The resulting conventional muon and neutrino fluxes (from $\pi$
and $K$ decays) agree well with earlier calculations, whereas our prompt
fluxes from charm decays are significantly lower than earlier estimates. 
Charm production is mainly considered as a well defined perturbative QCD 
process, but we also investigate a hypothetical non-perturbative intrinsic  
charm component in the proton. The lower charm rate implies better
prospects for detecting very high energy neutrinos from cosmic sources.
\end{quotation}
\section{Introduction}
The flux of muons and neutrinos at the earth has an important contribution from
decays of particles produced through the interaction of cosmic rays in the
atmosphere (for a recent introduction see \cite{Gaisser90}). This has an
interest in its own right, since it reflects primary interactions at energies
that can by far exceed the highest available accelerator energies. It is also a
background in studies of neutrinos from cosmic sources as attempted in large
neutrino telescopes, such as {\sc Amanda} \cite{amanda}, {\sc Baikal}
\cite{baikal}, {\sc Dumand} \cite{dumand} and {\sc Nestor} \cite{nestor}.

Here we present a detailed study of muon and neutrino  production in cosmic ray
interactions with nuclei in the  atmosphere using  Monte Carlo simulations
\cite{thunman}.

At GeV energies the atmospheric muon and neutrino fluxes are dominated by
`conventional' sources, \ie decays of relatively long-lived particles such as
$\pi$ and $K$ mesons. This is well understood from earlier studies
\cite{Volkova80,GSB88,Lipari93},  with which our investigations agree.  With
increasing energy, the probability increases that such particles interact in
the atmosphere before decaying. This implies that even a small fraction of
short-lived particles can give the dominant contribution to high energy muon
and neutrino fluxes. These `prompt' muons and neutrinos arise through
semi-leptonic decays of hadrons containing heavy quarks, most notably charm.

Available data in the multi-TeV energy range, obtained with 
surface and underground detectors (see {\it e.g.}
refs.~[10--13]), 
are still too discrepant to draw definitive conclusions on
the flux of prompt muons and neutrinos from charm. 
Furthermore, estimates of these prompt fluxes  
[7,14--21]
vary by a few orders of
magnitude due to the different models used to calculate the charm 
hadron cross section and energy spectra. 
This huge model dependence is due to the need of extrapolating charm
production data obtained at accelerator energies to the
orders-of-magnitude higher energies of the relevant cosmic ray collisions.  
Obviously, this extrapolation can only be trustworthy if starting from
proper charm production data and using a sound physical model. 
The main contribution of our study is in this context. First, we use 
recent charm production cross section measurements that form a consistent
set of data, but disagree with some of the early measurements that
were substantially higher. Secondly, we apply state-of-the-art models to 
simulate charm particle production through perturbative QCD processes 
in high energy hadron-hadron interactions. In addition, we investigate
a possible non-perturbative mechanism using the hypothesis of an 
intrinsic charm quark component in the nucleon. 

In the following we first (section 2) discuss the generalities of cosmic ray
interactions in the atmosphere resulting in a set of transport or cascade
equations for particle propagation. These equations are then solved by two
different methods: a direct Monte Carlo simulation of the cascade  interactions
(section 3) and a semi-analytic method (section 4) giving  consistent results
for the conventional and prompt muon and neutrino fluxes. Section 5 gives an
account of the Monte Carlo model used to  obtain the energy spectra of
secondaries in the basic hadron-hadron  interaction, in particular concerning
charm production in perturbative QCD.  In section 6 we investigate consequences
of the non-standard hypothesis  of a non-perturbative intrinsic charm quark
component in the nucleon. We then (section 7) compare our results with previous
model calculations  and discuss differences in terms of the different charm
production models.  We conclude (section 8) by some remarks and by putting our
results in a general  context of various astrophysical sources of high energy
neutrinos.

%
\section{Cosmic ray interactions in the atmosphere}
\subsection{The spectrum of cosmic rays}
Fluxes of secondary particles (hadrons and leptons) originate  from
nucleon--nucleon encounters, even when the nucleons are bound in nuclei,
because nuclear binding energies are much lower than the energies of interest
in this study (100\,GeV -- 10$^9$\,GeV). So the relevant quantity to consider
is the flux of nucleons. Following \cite{Gaisser90,Lipari93,Volkova87} we have
assumed a power law primary nucleon flux  
\begin{equation}  \label{eq:primary} 
\phi_{N}(E) \> \left[ \frac{\mbox{nucleons}}{\mbox{cm$^{2}$\,s\,sr\,GeV/$A$}}
\right] = \left\{ \begin{array}{ll} 1.7\,E^{-2.7} & \mbox{for}\,
E<5\cdot10^6\,\mbox{GeV}  \\ & \\  174\,E^{-3} & \mbox{for}\,
E>5\cdot10^6\,\mbox{GeV}. \end{array} \right. \end{equation} 
The normalisation constant 1.7 is derived \cite{Pal92} from the directly 
measured primary spectrum using balloon-borne emulsion chambers in JACEE
\cite{JACEE}. To within some 10\% this agrees with more indirectly derived
spectra based on measured atmospheric muon fluxes \cite{MACRO}, and is also
compatible with the data discussed in ref.~\cite{Honda}. The cosmic ray
composition is dominated by protons with only a smaller  component of neutrons
in nuclei \cite{Pal92,Honda}.  Only primary protons are considered here, since
in this study we are interested in quantities that are essentially independent
of the cosmic ray composition. At the energies of interest ($E \gsim
100$\,GeV), the cosmic ray flux can be considered isotropic (the anisotropy
being $\lsim 5\%$ \cite{Gregory82}).

\subsection{The model for the atmosphere} \label{sec:atmosphere}

In studying the propagation of particles through the atmosphere, an important
quantity is the amount of atmosphere $X$, in g/cm$^2$, traversed by the
particle. This so-called slant depth is the integral of the atmospheric
density from the top of the atmosphere downward along the trajectory of the
incident particle.  At distance $\ell$ from the ground along a direction at an
angle $\theta$ from the zenith, the slant depth is defined as
\begin{equation}\label{slantdepth} 
X(\ell ,\theta ) = \int_\ell^\infty \rho[h(l,\theta )] d l ,
\end{equation} 
where $\rho[h(l,\theta)]$ is the atmospheric density at the
altitude $h(l,\theta)$, \begin{equation} \label{eq:height} h(l,\theta) = \sqrt{
R_\earth^2 + 2 l R_\earth \cos \theta + l^2 } - R_\earth \,
\approx\,l\/\cos\theta\ +\ \frac{l^{2}}{2\,R_{\oplus}}\,\sin^{2}\theta.
\end{equation} Here $R_\earth$ is the radius of the Earth and the approximate
equality applies for zenith angles not too far from vertical, $\theta \lsim 
60^\circ$.

The atmospheric density is not a simple function of height, even neglecting
local atmospheric turbulence.  The temperature, which is related to the density
through the equation of state, decreases with increasing height until the
tropopause (8--17\,km), stays almost constant in the lower stratosphere
($-56.42^\circ$C up to 20--30\,km), then increases until the stratopause 
(50\,km) before decreasing again at the highest altitudes ($>50$ km). However,
since most particle interactions occur at heights between 10 and 40\,km
(demonstrated in Fig.\,\ref{fig:1} below), we need only a simple model for the
density profile of the stratosphere. We therefore adopt an isothermal model,
\begin{equation} \label{expatm} \rho(h) = \rho_0 e^{-h/h_0} , \end{equation}
with scale height $h_0=6.4$ km and $X_0=\rho_0 h_0=1300$\,g/cm$^2$, values
which adequately describe the density of the stratosphere ($<\!2\%$ error in
the vertical depth between 10 and 30\,km and $<\!16 \%$ between 30 and 40\,km).

Concerning the atmospheric composition, a good approximation, valid up to a
height of 100\,km, is 78.4\% nitrogen, 21.1\% oxygen and 0.5\% argon (obtained
from data in \cite{Allen83}). This leads to an average atomic number of
$\langle A \rangle = 14.5 $.


\subsection{Particle interactions with air nuclei} \label{sec:partint}

To obtain the flux of atmospheric muons and neutrinos one needs to consider the
particle production  mechanisms in strong interaction dynamics.    The cosmic
ray particles, represented by protons (see section 2.1),  interact with nuclei
in the atmosphere to produce  secondary particles. These proton-nucleus
collisions can, for our purposes,  be well represented by the simpler
proton-nucleon collisions  and a rescaling of the cross section 
\begin{equation} \sigma(pA) = A^{\alpha} \sigma(pN) \end{equation}  using a
power dependence on the number $A$ of nucleons in the target nucleus.  For
inclusive cross sections, with $\sigma(pN)$ of order 10\,mb,  the interaction
occurs with nucleons at the surface resulting in $\alpha\simeq 2/3$  as
verified experimentally.

The inelastic $pN$ interaction produces secondary hadrons with  a multiplicity
increasing essentially logarithmically with the cms energy 
($s_{pN}=2m_Nc^2E_p$). The formation time of a hadron is the normal strong 
interaction time scale. In the particles rest system this corresponds to  a
formation length of $\sim 1$\,fm,  which is Lorentz transformed with a
$\gamma$-factor to the target nucleus rest frame and thus becomes proportional
to the energy of the particle \cite{formlength}. Therefore, fast particles have
formation lengths that exceed the size of the  nucleus whereas only slow
particles are formed and can re-interact within the target nucleus. Therefore,
intra-nuclear cascade effects are not  important for the energetic particle
production studied here.
 
Of importance for our considerations are the energy distributions of  secondary
hadrons produced in the collisions 
\begin{equation} \label{eq:dn/de} \frac{\d
n(kA\to
hY;E_k,E_h)}{\d E_h} =  \frac{1}{\sigma_{kA}(E_k)} \frac{\d \sigma(kA\to
hY;E_k,E_h)}{\d E_h}, \end{equation}  
where  $ \d n(kA\to hY;E_k,E_h) $ is the
number of hadrons $h$ with energies between $E_h$ and $E_h + \d E_h$ produced
in the collision of the  incoming particle $k$ with an air nucleus of  atomic
number $A$, and $ \sigma_{kA} $ is the total inelastic cross section for
particle
$k$ -- nucleus $A$ collisions. Experiments studying proton--nucleus
\cite{helios89} and heavy ion \cite{Hoang94} collisions obtain energy spectra
that are approximately the same as in proton--proton collisions, confirming
that the interactions are essentially proton--nucleon. So we adopt an energy
distribution $ \d n_{kh} / \d E_h $ independent of the atomic number of the
target.

The energy spectra in Eq.\,(\ref{eq:dn/de}) can also be expressed as
\begin{equation} \label{eq:XFDISTR}
\frac{\d n_{kh}}{\d x_F} =
\frac{1}{\sigma_{kA}} \frac{\d \sigma(kA\to hY)}{\d x_F}
\end{equation}
in terms of the scaled longitudinal momentum or 
Feynman-$x$ variable $x_F=p_z/p_{z,max}$ ($\approx E_h/E_k$ at large 
energies) where the $z$-axis is along the incoming particle momentum.
If these distributions are independent of the cms energy (\ie incoming particle
energy $E_k$), then `Feynman scaling' holds. The validity or breaking of this 
scaling in different models for particle production is an important issue 
as will be demonstrated later. 

To obtain the energy spectra of the particles produced in proton-nucleon
collisions we use the Lund Monte Carlo simulation programs {\sc Pythia} and 
{\sc Jetset} \cite{pythia}. These have proven very successful in describing the
multi-particle final state in various kinds of interactions, including 
hadron-hadron collisions.  An advantage with this Monte Carlo approach is the
access to the complete  final state as well as a proper account of the decay of
unstable particles. Conventional muons and neutrinos are obtained from an
inclusive event sample generated with {\sc Pythia} in a mode simulating 
minimum bias proton-proton interactions (including diffractive scattering). The
particle production results from Lund model \cite{lund}  hadronization of
colour string fields between partons scattered in  semi-soft QCD interactions.
The prompt muons and neutrinos, on the other hand, are obtained from a
dedicated charm production simulation using {\sc Pythia}. Here, charm particles
arise from the hadronization of charm quarks produced in  the processes $gg\to
c\bar{c}$ and $q\bar{q}\to c\bar{c}$ as calculated with  leading order
perturbative QCD matrix elements. A more detailed account of the Monte Carlo
model is given in section 5. Since non-perturbative charm production  is
neither well established nor well defined, it is not part of our main  Monte
Carlo study but investigated separately  based on the intrinsic charm
hypothesis in section \ref{sec:IC}.


\subsection{Particle propagation in the atmosphere}
\label{sec:casceq}

Propagation of high energy particles through the atmosphere may be described by
a set of transport or cascade equations. In principle, the transport equations
for nucleons, mesons, unstable baryons and leptons are coupled,  but under the
reasonable  assumptions made below they can be greatly simplified.

Nucleons constitute the initial primary flux. We consider nucleon absorption
and regeneration in nucleon--air inelastic collisions, but neglect the
certainly small contribution to the nucleon flux from the interaction of
unstable hadrons with air nuclei. Absorption is described by the interaction
thickness $\lambda_{N}$ of nucleons $N$ in air, \ie the  average amount of
atmosphere (in g/cm$^2$)  traversed  between successive collisions with air
nuclei. It is given by
\begin{equation} \label{eq:intlength} \lambda_{N}(E) = { \rho(h) \over
\sum_A \sigma_{NA}(E) \, n_A(h) }, \end{equation}
where $n_A(h)$ is the number
density of air nuclei  of atomic number $A$ at height $h$ and $\sigma_{NA}(E)$
is the  inclusive inelastic cross section for collisions of nucleons with 
nuclei $A$. Note that to a good approximation $  \lambda_{N}(E) $
does not depend on the height $h$ because the atmospheric composition is
approximately independent of the height up to 100\,km.

Nucleon fluxes develop according to the cascade equation  
\begin{equation}
\label{eq:cn} \frac{\d \phi_{N}}{\d X} = - \frac{\phi_{N}} {\lambda_{N}} + 
S(NA\to NY) . \end{equation}  
Here $ \phi_{N}(E,X,\theta) $ is the nucleon
flux at slant depth $X$ in the atmosphere at zenith angle $\theta$, $
\lambda_{N}(E) $ has been defined in Eq.\,(\ref{eq:intlength}), and $ S(NA\to
NY) $ is the nucleon--nucleon regeneration function in air 
\begin{equation}
\label{eq:snn} S(NA\to NY) = \int_{E}^{\infty} \d E' \, \frac{\phi_{N}(E')}
{\lambda_{N}(E')}  \, \frac{\d n(NA\to NY;E',E)}{\d E} . \end{equation}

Mesons and unstable baryons, in addition to interact with the atmosphere, can
also decay. The decay length $ d(E) $, \ie the distance traveled in a mean
decay time, is simply \begin{equation} \label{eq:declength} d(E) = c \, \beta
\, \gamma \, \tau_ , \end{equation}  where $\tau$ is the particle proper
lifetime, $ \gamma =  E/m c^2 $ is its Lorentz factor, $ m$ its mass, and
$\beta$ its speed in units of the speed of light $c$.  The decay length
increases with particle energy because of relativistic time dilation; faster
particles can travel longer before decaying. This implies an increased
probability to interact before decaying. It is exactly because of this
energy-dependent competition between decay and interaction that the muon and
neutrino fluxes from charm mesons overcome those from pions and kaons at high
enough energy.

We assume that mesons and unstable baryons (collectively unstable hadrons) are
generated in nucleon--air collisions and regenerated in hadron--air collisions,
but neglect generation of unstable hadrons of other types in collisions of 
hadrons against air nuclei. This approximation is reasonable since the fluxes
of unstable hadrons are at least a factor of $\sim$\,10 smaller than the fluxes
of nucleons. Thus for mesons and unstable baryons we have
\begin{equation} \frac{\d
\phi_{M}}{\d X} = S(NA\to MY) - \frac{\phi_{M}} {\rho d_{M}} - \frac{\phi_{M}}
{\lambda_{M}} +  S(MA\to MY) , \end{equation} 
where $ \lambda_M(E) $ is the hadron interaction thickness in air (analogous to
Eq.\,(8)), and $ S(NA\to MY) $ and $ S(MA\to MY) $ are defined analogously to
Eq.\,(\ref{eq:snn}).

Finally we consider muons, muon-neutrinos and electron-neutrinos. At the
energies we are interested in, energy loss, absorption and muon decay can be
neglected and the transport equation for lepton $\ell$ = $\mu^{\pm}$,
$\nu_{\mu}$, $\bar{\nu}_{\mu}$, $\nu_{e}$, $\bar{\nu}_{e}$ contains only source
terms 
\begin{equation}
\label{eq:cl}
\frac{\d \phi_{\ell}}{\d X} = \sum_{M} S(M\to \ell Y)
, \end{equation} where the sum runs over all mesons and unstable baryons
decaying into muons and neutrinos $M$ = $\pi^{\pm}$, $K^{\pm}$, $K^0$,
$D^{\pm}$, $D^0$, $\bar{D}^0$, $D^{\pm}_{s}$, $\Lambda^{\pm}_c$. $ S(M\to
\ell Y) $ describes lepton production in hadron decays, 
\begin{equation} S(M\to
\ell Y) = \int_{E}^{\infty} \d E_M \, \frac{\phi_{M}(E_M)} {\rho d_{M}(E_M)}  \,
\frac{\d n(M\to \ell Y;E_M,E)}{\d E} , \end{equation} where $ \d n(M\to \ell
Y;E_M,E) $ is the number of leptons with energy between $ E $ and $E + \d E $
produced in the decay of hadron $M$.

Muons and neutrinos born out of pions and kaons are traditionally called
`conventional,' while those born out of charmed hadrons are called `prompt.'
This originates from the fact that up to \mbox{$\sim10^7$}\,GeV  very
short-lived charmed particles have negligible probability of being absorbed in
the atmosphere before decaying. Up to \mbox{$\sim10^7$}\,GeV the prompt flux is
therefore essentially independent of the zenith angle. The conventional muon
and neutrino fluxes are instead lower in the vertical direction, where the
amount of atmosphere traversed in a given meson decay length is larger. The
prompt flux is therefore relatively more important in the vertical direction,
and we will predominantly consider this direction.

\section{Simulation of cascade interactions}

One way of solving the transport equations described in the previous section
is to simulate the particle cascade with a Monte Carlo program. 
Here we describe our simulation algorithm. 

A cosmic ray proton is generated. Its energy is drawn from a flat distribution
in log\,$E$, and a weight is assigned to it in order to reproduce the shape of
the primary spectrum.

\begin{figure}[b]
\begin{center}
\includegraphics[width=7cm]{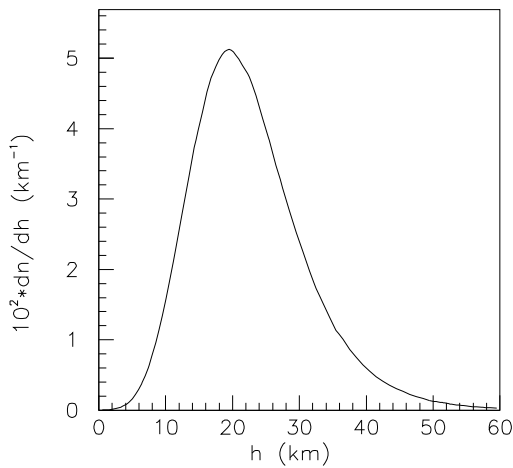}
\caption[junk]{{\it Distribution of the altitude for the 
primary interactions as obtained in the cascade simulations.}}
\label{fig:1} 
\end{center}
\end{figure}

An interaction height $h$ for the cosmic ray proton is then chosen in the
following way. Primary nucleons propagate down through the atmosphere according
to Eq.\,(\ref{eq:cn})  without the regeneration term $S(NA\to NY)$. From the
solution to this equation, $ \phi(h) = \phi_{\infty} \, \exp\left(
-X(h)/\lambda_N \right) $, the probability distribution for the primary
interaction height can be obtained. Using a standard Monte Carlo technique, we
generate this distribution by replacing $ \phi(h)/\phi_{\infty} $ with a
uniform random number  $ R \in \, ]0,1[$ and then solving for the interaction
height $h$. This can be done analytically for the isothermal atmospheric model
in Eq.\,(4) in the vertical direction, for which we obtain \begin{equation}
\label{eq:intheight}  h = - h_{0} \, \ln \left( \frac {-\lambda_{N}\,\ln\,R}
{X_{0}} \right) \, . \end{equation} 
Fig.\,\ref{fig:1} shows the height distribution so obtained (neglecting the
logarithmic energy dependence of $\lambda_N$), which confirms 
that, under the assumptions made, most particle interactions occur at heights
between 10 and 40\,km.

\begin{figure}[b]
\begin{center} 
\includegraphics[width=16cm]{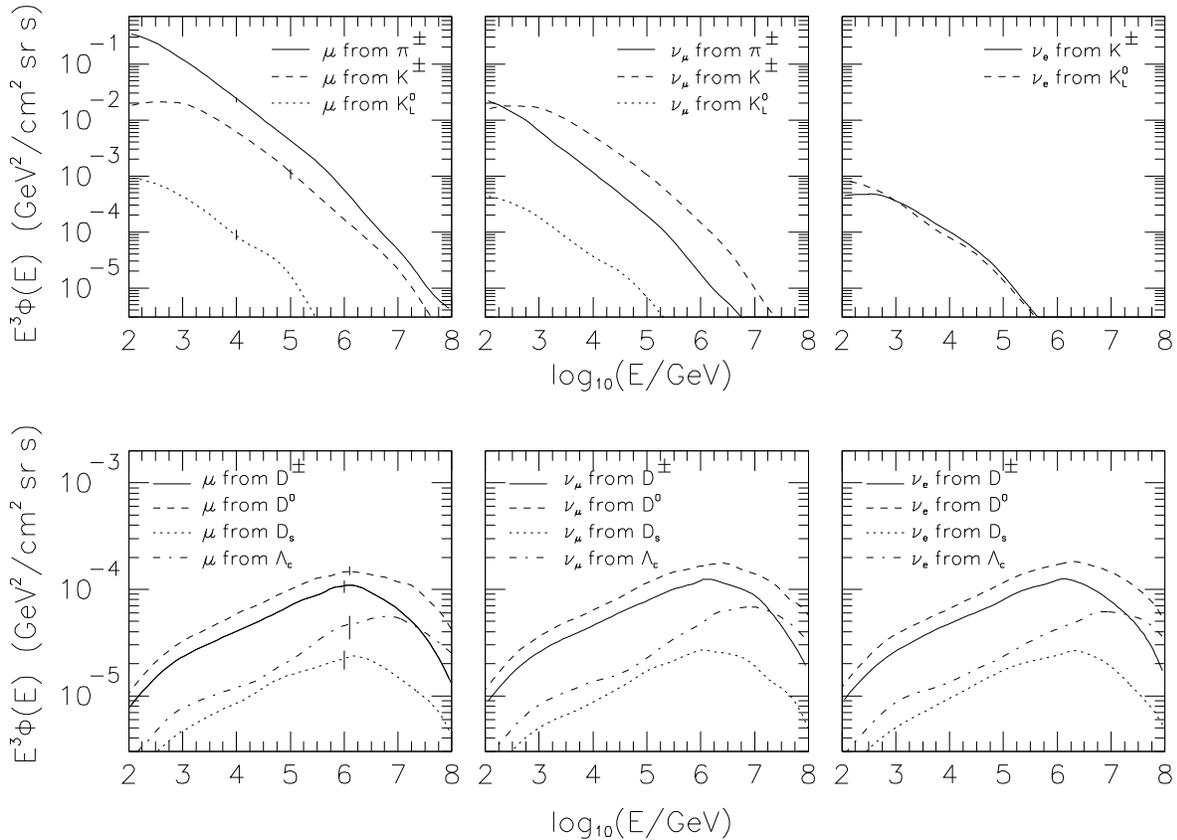}
\caption[junk]{{\it The $E^3$-weighted flux of muons ($\mu^+ +\mu^-$), 
muon-neutrinos ($\nu_{\mu} + \bar{\nu}_{\mu}$) and 
electron-neutrinos ($\nu_e + \bar{\nu}_e$) from decays of the 
specified particles. The error bars indicate the statistical precision 
of the Monte Carlo simulation.}}
\label{fig:2} 
\end{center} 
\end{figure}

A proton-nucleon interaction is then generated in full detail with  {\sc
Pythia} \cite{pythia} resulting in a complete final state of particles.
Secondary particles are followed through the atmosphere where they decay 
or interact producing cascades. 
Secondary nucleons give a flux that is rather small compared to the primary
flux 
and could therefore be neglected as a first approximation. To be more precise,
we do include the main effect of this correction by taking into account 
secondary nucleons that have an energy of at least 30\% of the primary one.
Nucleons with a lower energy give a negligible contribution compared to the
primary flux due to its steep energy spectrum Eq.~(\ref{eq:primary}). 
These leading nucleons emerging in the interactions are therefore allowed 
to generate a secondary interaction at a height 
\begin{equation} 
\label{eq:secint}
h = -h_{0} \,
\ln\left(e^{-H/h_0}-\frac{\lambda_{N}\,\ln\,R'} {X_{0}} \right), 
\end{equation}
obtained analogously to Eq.\,(\ref{eq:intheight}) but taking into account 
the finite height $H$ of the primary interaction. The procedure is iterated
until the energy of the leading nucleon from an interaction falls below
30\% of the primary cosmic ray proton energy.

Secondary mesons and unstable baryons are traced through the atmosphere until
they either decay or interact. Which of these occurs is decided by comparing
simulated decay and interaction lengths \begin{equation}
L_{dec}=-d_M(E)\,\ln\,R_1 \end{equation} and  \begin{equation}
\label{eq:intlen} L_{int} = H + h_{0} \,
\ln\left(e^{-H/h_0}-\frac{\lambda_{M}\,\ln\,R_2} {X_{0}} \right), 
\end{equation} where $R_1$ and $R_2$ denote uniform random numbers $ \in \,
]0,1[$ and $H$ is the height at which the traced particle has been produced.
The decay length $d_M(E)$ and the interaction thickness $\lambda_M$ are given
in Eqs.\,(\ref{eq:declength}) and~(\ref{eq:intlength}) respectively, the
atmospheric scale height $h_0$ and depth $X_0$ are defined in sect.~2.2.
Eq.\,(\ref{eq:intlen}) is obtained in a way analogous to Eq.\,(\ref{eq:secint}).

Particle decays are fully simulated with daughter particle momenta. In case of
interactions, the interacting particle is  regenerated in the same direction
but with degraded energy, chosen according to the appropriate leading particle
spectrum. Considering only the most energetic `leading' particles in secondary 
interactions is justified  because they give the dominant contribution to the
lepton fluxes. Moreover, other particles with lower energy are much fewer than
the particles of the same type and energy produced in  primary interactions.

\begin{figure}[b]
\begin{center}
\includegraphics[width=16cm]{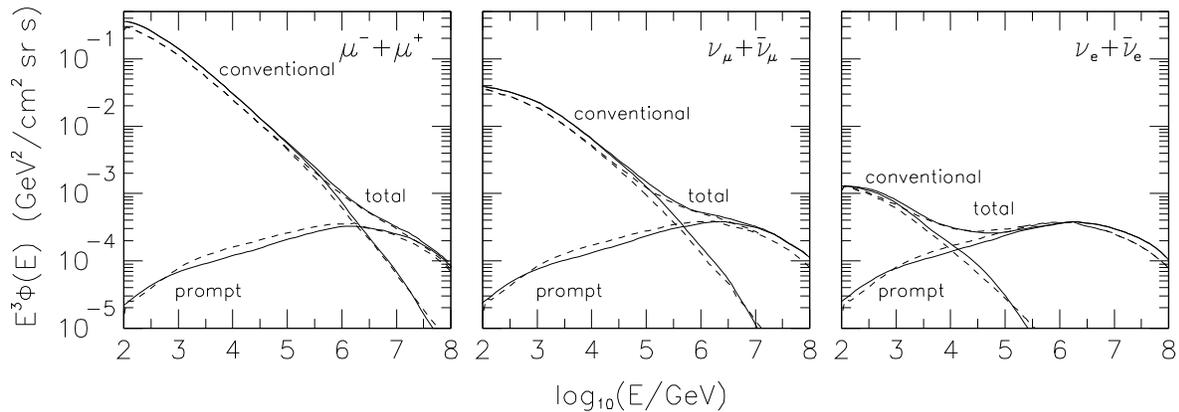}
\caption[junk]{{\it The $E^3$-weighted vertical flux of muons, muon-neutrinos 
and electron-neutrinos from conventional ($\pi , K$ decays) and prompt (charm
decays) sources and their sum (`total').
The solid lines are from the cascade simulation (section 3)
and the dashed lines are from the analytic $Z$-moment method (section 4).}}
\label{fig:3} 
\end{center}
\end{figure}

The particle decay--interaction chain is then repeated until all particles have 
decayed, have hit the ground or their energy has fallen below the minimum
energy of interest, $100$\,GeV.  Energy spectra for muons and neutrinos are
finally obtained by counting muons and neutrinos with the initially-assigned
primary proton weight.

The resulting fluxes of muons and neutrinos from different parent particles 
are shown in Fig.\,\ref{fig:2}. For charmed particles the figure clearly
demonstrates the dominance of the $D^{\pm,0}$ mesons, while for conventional
fluxes the dominant source varies with the type of lepton considered. Summing
the various contributions gives the inclusive fluxes in Fig.\,\ref{fig:3}. As
can be seen, the prompt contribution from charmed particles dominates  at high
energies.

The results for the inclusive prompt and conventional fluxes can be
parametrised as (similarly to \cite{Volkova80}, cf.\ with the primary flux and
Eq.\,(\ref{eq:interpol}) below)  
\begin{equation} \label{eq:fiteq} \phi(E)=\left\{\begin{array}{ll}
N_{0}\,E^{-\gamma-1}/(1+A\,E)\,, & E<E_{0} , \\ & \\
N_{0}'\,E^{-\gamma'-1}/(1+A\,E) \,, & E>E_{0} , \\ \end{array} \right.
\end{equation} with an accuracy of typically better than 10\% \ using the 
fitted parameter values in Table~\ref{tab:fluxfit}.

\begin{table}
\begin{center}
\begin{tabular}{|l|llllll|}
\hline
                 & \multicolumn{1}{c}{$N_{0}$}
                 & \multicolumn{1}{c}{$\gamma$} 
                 & \multicolumn{1}{c}{$A$} 
                 & \multicolumn{1}{c}{$E_{0}$} 
                 & \multicolumn{1}{c}{$\gamma'$}
                 & \multicolumn{1}{c|}{$N_{0}'$} \\
\hline
\makebox[25mm][l]{Conventional} $\mu^{-}+\mu^{+}$    & $2.0\cdot 10^{-1}$ & 
$1.74$ & $7.0\cdot 10^{-3}$ & $5.3\cdot 10^{5}$ & $2.10$ & $2.2\cdot 10^{1}$ \\
\makebox[25mm][l]{Prompt} $\mu^{-}+\mu^{+}$   & $1.4\cdot 10^{-5}$ & $1.77$ & 
$2.8\cdot 10^{-8}$ & $9.2\cdot 10^{5}$ & $2.01$ & $4.3\cdot 10^{-4}$ \\ \hline
\makebox[25mm][l]{Conventional} $\nu_{\mu}+\bar{\nu}_{\mu}$  & 
$1.2\cdot 10^{-2}$ & $1.74$ & $2.0\cdot 10^{-3}$ & $6.3\cdot 10^{5}$ & $2.17$ & 
$3.8\cdot 10^{0}$ \\
\makebox[25mm][l]{Prompt} $\nu_{\mu}+\bar{\nu}_{\mu}$ & $1.5\cdot 10^{-5}$ & 
$1.77$ & $3.1\cdot 10^{-8}$ & $1.2\cdot 10^{6}$ & $1.99$ & $3.1\cdot 
10^{-4}$ \\ \hline
\makebox[25mm][l]{Conventional} $\nu_{e}+\bar{\nu}_{e}$    & $4.2\cdot 10^{-4}$ 
& $1.63$ & $7.0\cdot 10^{-3}$ & $5.0\cdot 10^{4}$ & $2.12$ & $8.4\cdot 
10^{-2}$ \\
\makebox[25mm][l]{Prompt} $\nu_{e}+\bar{\nu}_{e}$     & $1.5\cdot 10^{-5}$ & 
$1.77$ & $3.0\cdot 10^{-8}$ & $1.2\cdot 10^{6}$ & $2.02$ & $4.9\cdot 
10^{-4}$ \\ \hline
\end{tabular}
\end{center}
\caption{\em Values of parameters in Eq.\,(\protect\ref{eq:fiteq}) obtained
from fits to the Monte Carlo results of the cascade simulations in
 Fig.\,\protect\ref{fig:3}.}
\label{tab:fluxfit}
\end{table}

\section{Approximate analytic solutions}
\label{sec:zmom}

Approximate analytic expressions for the muon and neutrino fluxes can be found
from the cascade equations in sect.~\ref{sec:casceq} by interpolation of
high-energy and low-energy asymptotic solutions. This is done in the standard
treatment for power law primary spectra and scale-invariant interaction cross
sections \cite{Gaisser90,Volkova80,Lipari93}. We wish to generalize 
the standard treatment to include non-scaling effects. 

The cascade equations for nucleons and mesons (and unstable baryons)
can, using Eqs.~(9,10, 12), be written 
\begin{eqnarray} \frac{\d \phi_{N}}{\d X} & = & -
\frac{\phi_{N}} {\lambda_{N}} +  Z_{NN} \frac{\phi_{N}}{\lambda_{N}}, \\
\frac{\d \phi_{M}}{\d X} & = & - \frac{\phi_{M}} {\rho d_{M}} - \frac{\phi_{M}}
{\lambda_{M}} +  Z_{MM} \frac{\phi_{M}}{\lambda_{M}} + Z_{NM}
\frac{\phi_{N}}{\lambda_{N}}, 
\end{eqnarray} 
where the spectrum-weighted moments for generation $Z_{NM}$ and regeneration 
$Z_{NN}, Z_{MM}$ in hadronic collisions are generally defined as 
\begin{equation} \label{eq:Zkh-general} 
Z_{kh} = \int_{E}^{\infty} \d E' \, 
\frac{\phi_k(E',X,\theta)}{\phi_k(E,X,\theta)} \,
\frac{\lambda_{k}(E)} {\lambda_{k}(E')}  \, \frac{\d n(kA\to hY;E',E)}{\d E} .
\end{equation}
It is assumed that
the fluxes of nucleons, mesons, unstable baryons, muons and neutrinos can 
be approximated in the factorized form 
$\phi_i(E,X,\theta) = E^{-\beta_i} \, \phi_i(X,\theta) , $ with appropriate
values of the exponents $\beta_i$ 
in the low- and high-energy asymptotic regimes. 
We consider both a primary spectrum 
$ \phi_N(E) \propto E^{-\gamma-1} $
with a constant spectral index $\gamma$, and the
primary spectrum with a knee as given in Eq.\,(\ref{eq:primary}).  
Since nucleon and meson fluxes develop rapidly in the atmosphere, 
their ratios are essentially independent on the depth $X$. 
To a good approximation one 
therefore obtains the {\em energy-dependent} spectrum-weighted moments 
\begin{equation} \label{eq:zkh} 
Z_{kh}(E) =
\int_{E}^{\infty} \d E' \, \left( \frac{E'}{E} \right) ^ {-\gamma-1} \,
\frac{\lambda_{k}(E)} {\lambda_{k}(E')}  \, \frac{\d n(kA\to hY;E',E)}{\d E},
\end{equation}
which we will estimate numerically. In previous studies it has been 
assumed that Feynman scaling holds, such that the distributions in  
$ x_F = E_{h}/E_{k} $ of produced particles are energy independent (cf. 
Eq.~(\ref{eq:XFDISTR})). With this additional assumption the $Z$-moments 
become simply 
\begin{equation}\label{eq:Zscaling}  
Z_{kh}^{scaling} = \int_0^1 x_F^\gamma \, { \d n_{kh} \over \d x_F } \, \d x_F .
\end{equation} 
We will not make this scaling assumption, but investigate its validity 
by our calculation of the energy-dependent $Z$-moments.

An approximate solution for the nucleon fluxes is then 
\begin{equation}
\label{eq:phi_n}
\phi_{N} = e^{-X/\Lambda_N} \, \phi_{N}(E) , 
\end{equation}
 where the nucleon attenuation length $ \Lambda_N $ is defined as 
\begin{equation}
\label{eq:attenlength} 
\Lambda_N(E) = \frac{\lambda_{N}(E)} {1 - Z_{NN}(E)} ,
\end{equation} 
and $ \phi_{N}(E) $ is the primary nucleon spectrum.  

Concerning mesons and unstable baryons, at sufficiently low energies the
interaction term can be neglected and so also the regeneration term. One then
finds  
\begin{equation} \phi_{M}^{low} =  \frac{Z_{NM}}{1-Z_{NN}} \, \frac{\rho
d_M}{\Lambda_N} e^{-X/\Lambda_N} \,  \phi_{N}(E) .  \end{equation}  In the high
energy regime it is the decay term that can be neglected, and one finds in a
similar way \begin{equation}  \phi_{M}^{high} = \frac{Z_{NM}}{1-Z_{NN}} \,
\frac{ e^{-X/\Lambda_M} - e^{-X/\Lambda_N} } { 1 - \Lambda_N/\Lambda_M } \,
\phi_{N}(E) ,  \end{equation}  with the meson attenuation length $\Lambda_M $
defined analogously to Eq.\,(\ref{eq:attenlength}). Notice that at high energies
the spectral index of the meson flux is the same as that of the primary flux, $
\phi_M^{high} \propto E^{-\gamma-1} $, while at low energies the meson spectrum
is flatter by one power of energy, $ \phi_M^{low} \propto E^{-\gamma} $,
because of the implicit proportionality of the decay length $d_M$ to the energy
$E$.

The spectrum-weighted moments for hadron generation in hadron--air collisions 
can be rewritten as 
\begin{equation} \label{eq:zmom} Z_{kh}(E_h) =
\int_{E_h}^{\infty} \d E_k \, \frac {\sigma_{kA}(E_k)} {\sigma_{kA}(E_h)} \,
\left( \frac {E_k} {E_h} \right) ^ {-\gamma-1} \, \frac{\d n(kA\to
hY;E_k,E_h)}{\d E_h} . \end{equation}  
The numerical evaluation of these $Z$-moments was made by applying the 
prefactor in the integrand  to the hadron spectra  ${\d n_{kh}}/{\d E_h}$
generated at different incoming energies $E_k$ between $10^2$ and $10^9$\,GeV
with the {\sc Pythia} Monte Carlo (see section 5 for details on the generation
mechanism). Total inelastic cross sections $\sigma_{kA}(E)$ in
Eq.\,(\ref{eq:zmom})  were taken from ref.~\cite{pdb} when available. The
spectra of regenerated kaons and $D$ mesons,  which cannot be used as beam
particles in {\sc Pythia} simulations,  were approximated by the leading pion
spectrum obtained in pion-proton collisions. Regeneration of
$\Lambda_c$-baryons was mimicked using an ordinary $\Lambda$ as a beam particle
in {\sc Pythia} and extracting the spectrum of leading $\Lambda$'s.  The
resulting muon and neutrino spectra are rather insensitive to these
approximations, since they are slowly varying functions of kaon and heavier
hadron regeneration $Z$-moments (they enter only through the combination $A_M$
in Eq.\,(\ref{eq:AM}) below).

\begin{figure}[t]
\begin{center}
\includegraphics[width=16cm]{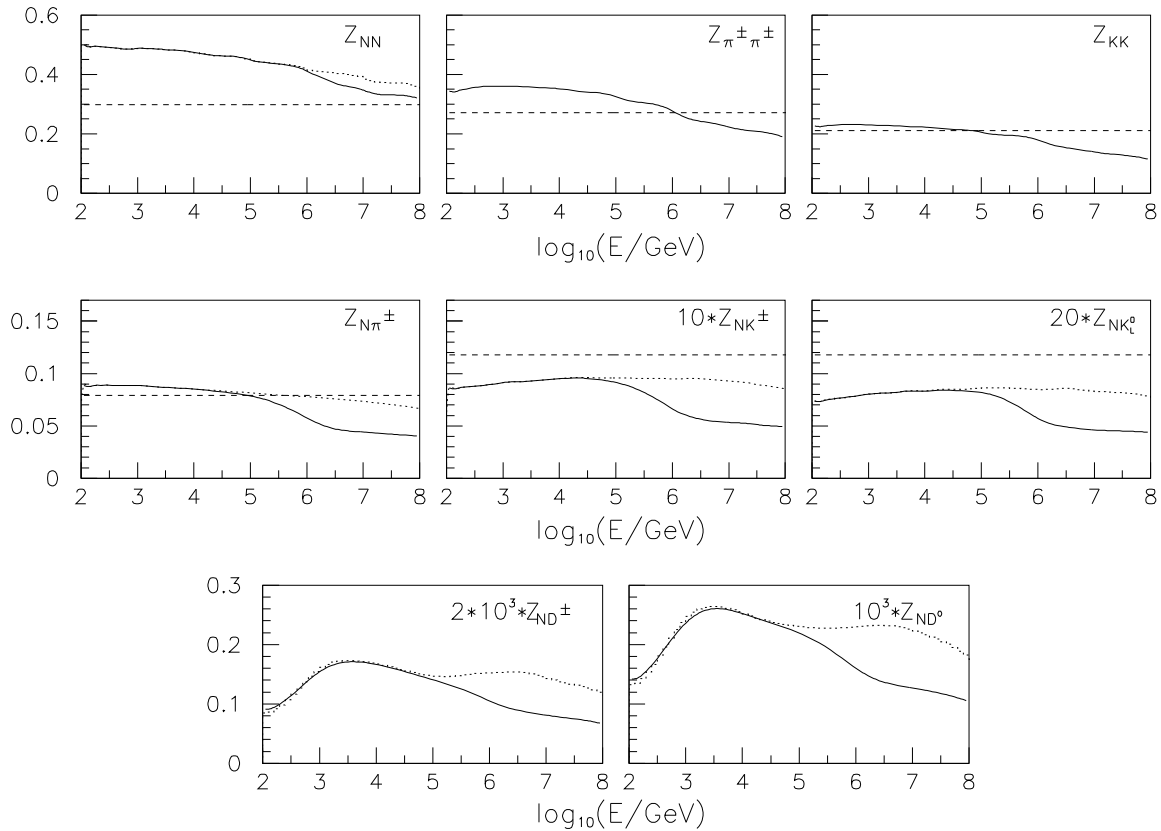}
\caption[junk]{{\it Energy-dependence of 
production $Z_{kh}$-moments, Eq.\,(\protect\ref{eq:zmom}), 
for incoming particle $k$ producing hadron $h$. 
Solid lines are the results of our model using the initial spectrum
 with a `knee', Eq.\,(\protect\ref{eq:primary}), whereas the dotted lines are
obtained 
with a constant spectral index $\gamma=1.7$. Dashed lines show 
the values of \cite{Lipari93} based on Feynman scaling.}}
\label{fig:zmom} 
\end{center}
\end{figure}

For the charm $Z$-moments one has to consider that  the charm cross section
need not scale with the target atomic number $A$  in the same way as the total
inelastic cross section. The ratio of the $A^{1}$-dependence in our charm cross
section  (see section 5.2) and the inclusive $A^{2/3}$-dependence (see section
2.3) gives a $A^{1/3}$-dependence in the $Z$-moments for charm, which are
included  in our results (\eg  Fig.~\ref{fig:zmom}).

The energy dependence of the hadron generation and regeneration $Z$-moments is
shown in Fig.\,\ref{fig:zmom} for a constant spectral index $\gamma=1.7$ 
(dotted lines) and for a primary spectrum with a knee as in
Eq.\,(\ref{eq:primary}) (solid lines). For comparison, we also show the
constant values given by Lipari~\cite{Lipari93} under the assumptions of
energy-independent inelastic cross sections and Feynman scaling of the meson
spectra \cite{Frazer72,Garraffo73}, i.e. with a $Z$-moment defined by 
Eq.~(\ref{eq:Zscaling}). It is clear from the figures that variations of the
$Z$-moments with energy are non-negligible, in particular when the changing
slope ($\gamma$) of the primary spectrum is included. On the other hand, this
energy dependence of the $Z$-moments, and of the interaction lengths $\lambda$,
is mild with respect to the rapid decrease of the primary flux with increasing
energy. For example, differentiation of Eq.~(\ref{eq:phi_n}) gives
\begin{equation}
{E\over\phi_N} {{\rm d}\phi_N\over {\rm d}E} =
{X\over\Lambda_N} \, {E\over\Lambda_N} {{\rm d}\Lambda_N\over {\rm d}E} 
- (\gamma+1),
\end{equation}
where the first term comes from the energy dependence of $Z$ and $\lambda$ and
the second term from the primary spectrum. Numerically, the former is only 0.1,
much smaller than the latter which is 2.7 or 3.0. This demonstrates that the
treatment of non-scaling effects as  corrections embodied through
energy-dependent $Z$-moments is justified  irrespectively of the derivation
leading to Eq.\,(\ref{eq:zkh}).

\begin{table}[tb]
\begin{center}
\begin{tabular}{|l|cccccccc|}
\hline
Particle & $\pi^{\pm}$ & $K^{\pm}$ & $K_{L}$ & $D^{\pm}$ & $D^{0}$ & 
$\Lambda_{c}$ & $D_{s}$ & $J/\psi$ \\ \hline
$\varepsilon$ [GeV] & 115 & 850 & 205 & $3.74\cdot10^{7}$ & $9.47\cdot10^{7}$ & 
$2.57\cdot10^{8}$ & $9.33\cdot10^{7}$ & $ 8.8\cdot10^{15}$ \\ 
\hline
\end{tabular}
\end{center}
\caption{\em Critical energies for hadrons traversing the atmosphere 
in the vertical direction.}
\label{tab:epsilon} 
\end{table}

We can now find approximate asymptotic solutions for the muon and neutrino
fluxes. In the asymptotic regimes meson fluxes are well approximated by power
laws, $ \phi_M(E) \propto E^{-\beta}$,  with $\beta=\gamma $ in the low
energy case and $ \beta=\gamma+1 $ in the high energy case. Hence the
source terms in the lepton cascade equation (Eq.\,\ref{eq:cl})) can be rewritten
as \begin{equation} S(M\to \ell Y) =  Z_{M\to \ell,\beta+1} \frac{\phi_M}{\rho
d_M} , \end{equation} 
with  decay spectrum-weighted moments defined
by\footnote{To keep the analogy with the (re)generation $Z$-moments, which
include the multiplicity of the final state,  we include the branching ratio
$\mbox{\rm Br} (M\to \ell X) $ into the definition of the decay $Z$-moments.
This differs from \cite{Lipari93}.}  
\begin{equation} Z_{M\to \ell,\beta+1} =
\int_{E}^{\infty} \d E_M \, \left( \frac{E_M}{E} \right) ^ {-\beta} \,
\frac{d_{M}(E)}{d_{M}(E_M)}   \, \frac{\d n(M\to \ell Y;E_M,E)}{\d E} .
\end{equation} 
Integrating the lepton cascade equations over the line of sight,
one then obtains the following expressions for the lepton fluxes deep in the
atmosphere ($X\to \infty$): \begin{equation} \phi_{\ell}^{low} = Z_{M\to
\ell,\gamma+1} \,  \frac{Z_{NM}}{1-Z_{NN}} \,  \phi_{N}(E) \end{equation} for
leptons coming from low energy mesons and 
\begin{equation} \label{eq:phihigh} \phi_{\ell}^{high}
=  Z_{M\to \ell,\gamma+2} \,  \frac{Z_{NM}}{1-Z_{NN}} \,  \frac{
\ln\left({\Lambda_M}/{\Lambda_N}\right) } {1 - \Lambda_N/\Lambda_M} \,
\frac{\varepsilon_M}{E} \, \phi_{N}(E) \end{equation}
for leptons coming from high energy mesons (with $E_M\,\gg\,m_M$).
In Eq.\,(\ref{eq:phihigh}), $\varepsilon_M$ is a critical
meson energy separating the low energy and the high energy regimes, \ie 
where the meson dominantly decays or interacts, respectively. 
It depends on the
atmospheric profile and in general on zenith angle. For the exponential
atmospheric profile in sect.~\ref{sec:atmosphere} and in the vertical
direction, one has \cite{Gaisser90} \begin{equation} \varepsilon_M = \frac{m_M
c^2 h_0} {c \tau_M} . \end{equation} 
In Table~\ref{tab:epsilon} we have 
collected 
the critical energies $\varepsilon_M$ used in this study. For not too
large zenith angles $\theta \lsim 60^\circ$, Eq.\,(\ref{eq:height}) leads to 
$\varepsilon_M \propto 1/\cos\theta$ and $\phi_\ell^{high}$ depends on the
zenith angle.

To resume, the lepton fluxes from meson $M$ have the same spectral index of the
primary flux, $ \phi_\ell^{low} \propto E^{-\gamma-1} $, and are independent of
zenith angle at energies smaller than the meson critical energy $ \varepsilon_M
$, while they are steeper by one power of energy and depend on zenith angle at
energies above $\varepsilon_M$, $ \phi_\ell^{high} \propto E^{-\gamma-2} /
\cos\theta $. We see that at the energies of interest to us, pions and kaons
are above their critical energy and so generate `conventional' muons and
neutrinos, while charmed mesons are below their critical energy ($\sim
10^7$\,GeV) and  give `prompt' muons and neutrinos.

\begin{table}[tb]
\begin{center}
\begin{tabular}{|l|llll|}  \hline
   &  $\beta=1.7$  &  $\beta=2$  &  $\beta=2.7$ &  $\beta=3$ \\ 
\hline
$ \pi^{\pm}\rightarrow\mu^{-}/\mu^{+}$           &
  0.675  &  0.634   &  0.553   &  0.523 \\ 
$K^{\pm}\rightarrow\mu^{-}/\mu^{+}$             &
 0.253  &  0.227   &  0.183   &  0.169  \\ 
$K^{0}_{L}\rightarrow\mu^{-}/\mu^{+}$           &
 0.0543  &  0.0435   &  0.0273   &  0.0227 \\ 
$D^{\pm}\rightarrow\mu^{-}/\mu^{+}$             &
 0.0171 &  0.0124  &  0.00636  &  0.00490 \\ 
$D^{0}\rightarrow\mu^{-}/\mu^{+}$               &
 0.00880  &  0.00651  &  0.00346  &  0.00270 \\ 
$D_{s}\rightarrow\mu^{-}/\mu^{+}$               &
 0.00708 &  0.00516  &  0.00266  &  0.00205 \\ 
$\Lambda_{c}\rightarrow\mu^{-}/\mu^{+}$         &
 0.00377 &  0.00269  &  0.00130  &  0.000976 \\
$J/\psi\rightarrow\mu^{-}+\mu^{+}$         &
 0.045 &  0.040  &  0.033  &  0.030 \\ \hline
$\pi^{\pm}\rightarrow\nu_{\mu}/\bar{\nu_{\mu}}$ &
 0.0870 &  0.0607  &  0.0271  &  0.0194 \\ 
$K^{\pm}\rightarrow\nu_{\mu}/\bar{\nu_{\mu}}$   &
 0.221  &  0.196   &  0.153   &  0.139  \\
$K^{0}_{L}\rightarrow\nu_{\mu}/\bar{\nu_{\mu}}$ &
 0.0292  &  0.0216  &  0.0115  &  0.00894 \\ 
$D^{\pm}\rightarrow\nu_{\mu}/\bar{\nu_{\mu}}$   &
 0.0181 &  0.0134  &  0.00720  &  0.00566 \\ 
$D^{0}\rightarrow\nu_{\mu}/\bar{\nu_{\mu}}$     &
 0.00839  &  0.00636  &  0.00354  &  0.00283 \\ 
$D_{s}\rightarrow\nu_{\mu}/\bar{\nu_{\mu}}$     &
 0.00744 &  0.00550  &  0.00292  &  0.00228 \\ 
$\Lambda_{c}\rightarrow\nu_{\mu}/\bar{\nu_{\mu}}$ &
 0.00395 &  0.00284  &  0.00141  &  0.00107 \\ \hline 
$K^{\pm}\rightarrow\nu_{e}/\bar{\nu_{e}}$       &
 0.00653  &  0.00509   &  0.00298  &  0.00242 \\ 
$K^{0}_{L}\rightarrow\nu_{e}/\bar{\nu_{e}}$     &
 0.0517  &  0.0401   &  0.0235  &  0.0191 \\ 
$D^{\pm}\rightarrow\nu_{e}/\bar{\nu_{e}}$       &
 0.0187 &  0.0139  &  0.00756  &  0.00597 \\ 
$D^{0}\rightarrow\nu_{e}/\bar{\nu_{e}}$         &
 0.00870  &  0.00660  &  0.00372  &  0.00298 \\ 
$D_{s}\rightarrow\nu_{e}/\bar{\nu_{e}}$         &
 0.00767 &  0.00571  &  0.00306  &  0.00240 \\ 
$\Lambda_{c}\rightarrow\nu_{e}/\bar{\nu_{e}}$   &
 0.00404 &  0.00293  &  0.00148  &  0.00101 \\ \hline
\end{tabular}
\end{center}
\caption{\em Decay $Z$-moments $Z_{M\to \ell,\beta+1}$ for various decay
channels. (For the decay $J/\psi\rightarrow\mu^{-}+\mu^{+}$ a factor 2 is
included to account for both $\mu^-$ and $\mu^+$.)}
\label{tab:zdec}
\end{table}

The energy spectra of muons and neutrinos from decays of ultra-relativistic
mesons take a simple scaling form~\cite{Lipari93} 
\begin{equation} 
\d n(M\to\ell Y;E_M,E) = F_{M\to \ell} \! \left(\frac{E}{E_M}\right) \,
\frac{\d E}{E_M} , 
\end{equation}
 and the decay $Z$-moments are independent of energy
\begin{equation} 
Z_{M\to \ell,\beta+1} = \int_{0}^{1} \, \d x \, x ^ {\beta} \,
 F_{M\to \ell} \! \left( x \right) ,
\end{equation}  
with $ x = E/E_M$.
Approximate expressions for the
functions $ F_{M\to \ell} $ have been obtained for two and three body decay
channels in ref.~\cite{Lipari93}.\footnote{Because of our convention, the
functions $ F_{M\to \ell} $ in ref.~\cite{Lipari93} should be multiplied by the
branching ratio $\mbox{\rm Br}~(M\to~\ell~X) $.} Since there are many
semi-leptonic decay channels for charmed mesons and most of them have more than
three particles in the final state, we prefer to generate all decay spectra
within the Lund Monte Carlo. In Table~\ref{tab:zdec} we list the values of the
decay $Z$-moments for the spectral indices of interest in this study.

Finally we join the low and high energy solutions with the interpolation
\begin{equation}
\label{eq:interpol} \phi_\ell = \sum_{M} \frac{\phi_\ell^{low}
\phi_\ell^{high}}{\phi_\ell^{low} + \phi_\ell^{high}} =
\frac{\phi_{N}(E)}{1-Z_{NN}} \, \sum_M \frac{Z_{NM} Z_{M\to \ell,\gamma+1}} {1
\, + \, A_M E / \varepsilon_M} , \end{equation} with \begin{equation} 
\label{eq:AM} A_M =
\frac{Z_{M\to \ell,\gamma+1}} {Z_{M\to \ell,\gamma+2}} \, \frac{ 1 - \Lambda_N 
/ \Lambda_M} {\ln\left( \Lambda_M / \Lambda_N \right)} . \end{equation}

The fluxes of muons and neutrinos calculated according to
Eq.\,(\ref{eq:interpol}) using the previously-obtained energy-dependent
$Z$-moments are plotted in Fig.\,\ref{fig:3} as dashed lines.  It is satisfying
to see that the cascade simulations and the approximate analytic solutions, 
which are conceptually rather different, give results that are quite close. 
Detailed comparison of corresponding fluxes shows good agreement both for 
conventional and prompt muons and neutrinos. The differences are typically 
less than 20\% which is quite sufficient in this context. For the prompt 
leptons, this is below the uncertainty in our charm calculation (see section 5)
and far smaller than the differences between the different models
discussed in section \ref{sec:Comparison}.


\section{The model for particle production}
A model for particle production is needed to specify the energy spectra   of
secondaries in cosmic ray collisions with atmospheric nuclei.  As discussed in
section \ref{sec:partint}, collisions involving nuclei can be  reduced  to the
simpler proton-nucleon collision. This applies in particular when  only
energetic particles are of interest, as in our case.

The flux of conventional muons and neutrinos results from the decay of
relatively long-lived particles, such as $\pi$ and $K$ mesons. The production
of such hadrons, containing only light quarks ($u,d,s$), is dominated by
minimum bias proton-nucleon interactions (without large momentum transfers) and
receives a small contribution from diffractive interactions.  On the other
hand, the prompt muons and neutrinos arise through decays of  short-lived
particles, \ie dominantly charmed particles.  Charm quarks are, due to their
relatively large mass, usually considered to be produced in hard processes
which can be  described by perturbative QCD (pQCD). In the following, some
relevant details of the models implemented in the  {\sc Pythia} and {\sc
Jetset} Monte Carlo programs \cite{pythia}  will be discussed. The hypothetical
non-perturbative  intrinsic charm mechanism is discussed separately in
section~\ref{sec:IC}.
 
\subsection{Light particle production} 
The production of light hadrons is dominantly through minimum bias
hadron-hadron collisions. The strong interaction mechanism is here of a soft
non-perturbative nature that cannot be calculated based on proper theory, but
must be modelled. In the successful Lund model \cite{lund} hadron production
arise through the fragmentation of colour string fields between  partons
scattered in semi-soft QCD interactions \cite{pythia}. The essentially
one-dimensional colour field arising  between separated colour charges is
described by a one-dimensional  flux tube whose dynamics is taken as that of a
massless relativistic string.  Quark-antiquark pairs are produced from the
energy in the field through  a quantum mechanical tunneling process. The string
is thereby broken into smaller pieces with these new colour charges as
endpoints and, as the process is iterated, primary hadrons are formed. These
obtain limited momenta transverse to the string (given by a Gaussian of a few
hundred MeV width) but their longitudinal momentum may be large as  it is given
by a probability function in the fraction of the available  energy-momentum in
the string system taken by the hadron.   The iterative and stochastic nature of
the process is the basis for   the implementation of the model in the {\sc
Jetset} program \cite{pythia}.

A non-negligible contribution to the inclusive cross section is given by
diffractive interactions.  These are also modeled in {\sc Pythia} \cite{pythia}
using cross sections from a well functioning Regge-based approach and 
simulating the diffractively produced final state using an adaptation of  the
Lund string model. These diffractive events are included in our  simulations
and contribute rather less than 10\% to the final results. 

\subsection{Perturbative production of charm}
\label{sec:charm}
Charm production is strongly suppressed in hadronisation models that  are
commonly used in detailed comparisons with experimental data on  multihadron
production in various high energy collisions. The tunneling  mechanism in the
Lund model \cite{lund} gives a production probability of different quark
flavours as $u\bar{u} : d\bar{d} : s\bar{s} : c\bar{c} \simeq 1 : 1 : 0.3 :
10^{-11}$, \ie charm production in the hadronization phase can here be safely
neglected.

\begin{figure} 
\begin{center}
\includegraphics[width=16cm]{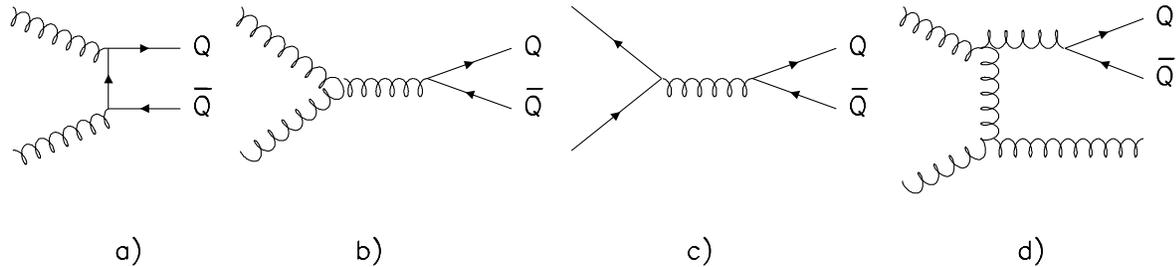}
\caption[junk]{{\it Illustration of charm production in pQCD. 
The leading order 
processes (a,b,c) and an important next-to-leading order process (d).}}
\label{fig:Feynman} 
\end{center}
\end{figure}

Charm quarks are instead considered to be produced in perturbative QCD
processes in accordance with the relatively large charm quark mass. To leading
order (LO) in the coupling constant, \ie $\cal{O}$$(\alpha_s^2)$, these are the
gluon-gluon fusion process $gg\to c\bar{c}$ and the quark-antiquark
annihilation process $q\bar{q}\to c\bar{c}$ as shown in
Fig.\,\ref{fig:Feynman}abc. The charm production cross section is calculated
using the usual convolution of parton densities $f_i$ in the colliding hadrons
and the hard parton level cross section $\hat{\sigma}$ from pQCD, \ie
\begin{equation} \label{SIGMACC}
\sigma = \int \int \int dx_1dx_2d\hat{t} \;\; f_1(x_1,Q^2)\; f_2(x_2,Q^2)\;  
         \frac{d\hat{\sigma}}{d\hat{t}}
\end{equation} 
Here, $x_i$ are the parton longitudinal momentum fractions in the hadrons and 
$\hat{t}$ is the Mandelstam momentum transfer at the parton level.  $Q^2$ is
the factorization scale defining at what momentum transfer  the parton
densities are probed and also regulating the amount of  pQCD scaling
violations; we have used $ Q^2=(m^2_{\bot c}\,+\,m^2_{\bot \overline{c}})/2$,
where $m^2_{\bot}=m^2\,+\,p^2_{\bot}$.  The charm quark mass introduces a
threshold in the invariant mass of the  parton level subsystem, \ie
$\hat{s}=x_1x_2s > 4m_{c}^{2} c^4$.  The dominating contribution to the cross
section comes from the region  close to this threshold, since $d\sigma
/d\hat{s}$ is a steeply falling distribution. It is therefore  important to use
QCD matrix elements with the charm quark mass explicitly  included. The
numerical value used is $m_{c}=1.35\,GeV/c^2$ together with 
$\Lambda_{QCD}=0.25\,GeV$ (in accordance with using the $MRS\,G$  parton
density parametrisation \cite{mrsg}).

Next-to-leading order (NLO), \ie $\cal{O}$$(\alpha_s^3)$, cross sections for 
heavy flavour production in hadron collisions have been calculated in  pQCD
\cite{Nason1,Nason2}.  Compared to the leading order results there is an
overall increase of the cross section of about a factor of two.  This does not
demonstrate a bad convergence of the perturbative series,  since the main NLO
contribution is associated with a process that does not  appear in leading
order charm production. This is the gluon scattering process $gg\to gg$, which
has a much larger cross section than the leading order charm processes and is
of a comparable magnitude when including the NLO correction  $g\to c\bar{c}$
shown in Fig.\,\ref{fig:Feynman}d.  Since the NLO distributions of the charm
quark transverse momentum and rapidity to a reasonable approximation have the
same shape as the LO ones, we take  the NLO results into account by rescaling
the cross section with an overall factor $K=2$.  Still higher order corrections
have not been calculated, but their effect should be significantly smaller
since there should be no such additional  process entering like for NLO. The
factor $K=2$ is indeed consistent  when comparing the leading order cross
sections with experimental results \cite{Cifarelli}.

Another estimate of higher order corrections can be obtained from charm
production in the simulated parton cascades implemented in {\sc Pythia}.  These
represent a leading logarithm approximation of multiple parton emission from
the incoming and scattered partons in basic QCD $2\to 2$ processes.  Charm
quark production arises here through the perturbative QCD gluon splitting
process $g\to c\bar{c}$, \ie basically the same as in the NLO matrix element
calculation. This approximate charm production to arbitrary order in pQCD gives
a contribution of the same magnitude as the LO matrix elements, thus confirming
the use of a renormalisation factor of $K=2$.  Furthermore, the energy spectra
of the charm quarks from this higher order treatment are very similar to those
from the LO calculation. If anything, they rather tend to be slightly softer
than the LO distribution \cite{thunman}.  Since it is the hardest part of the
spectrum that gives the largest contribution to the high energy neutrino and
muon spectra, one may conclude that taking the higher order corrections into
account through a global $K$-factor renormalisation and keeping the
leading-order shape of the charm quark energy spectra, as we have done, is
sufficient for the precision needed in this study.

\begin{figure}
\begin{center} 
\includegraphics[width=8cm]{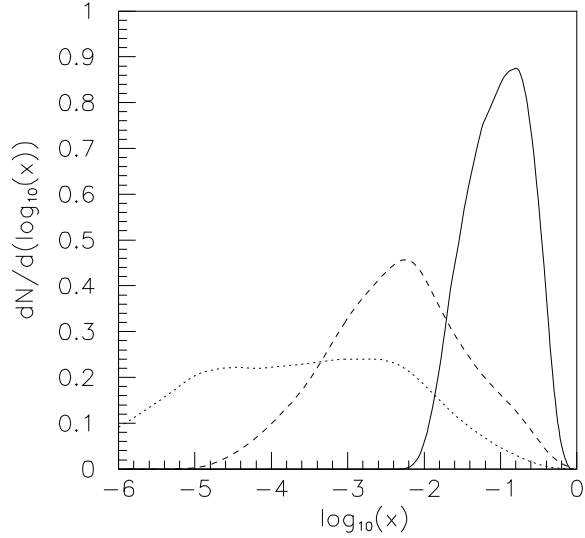}
\caption[junk]{{\it Distribution of momentum fraction $x$ for the initial 
partons 
entering the QCD charm production subprocesses. The curves represent  
three different beam particle energies: $10^{3}$\,GeV (solid), 
$10^{6}$\,GeV (dashed) and $10^{9}$\,GeV (dotted).}}
\label{fig:5} 
\end{center} 
\end{figure}

\begin{figure}[tb]
\begin{center} 
\includegraphics[width=14cm]{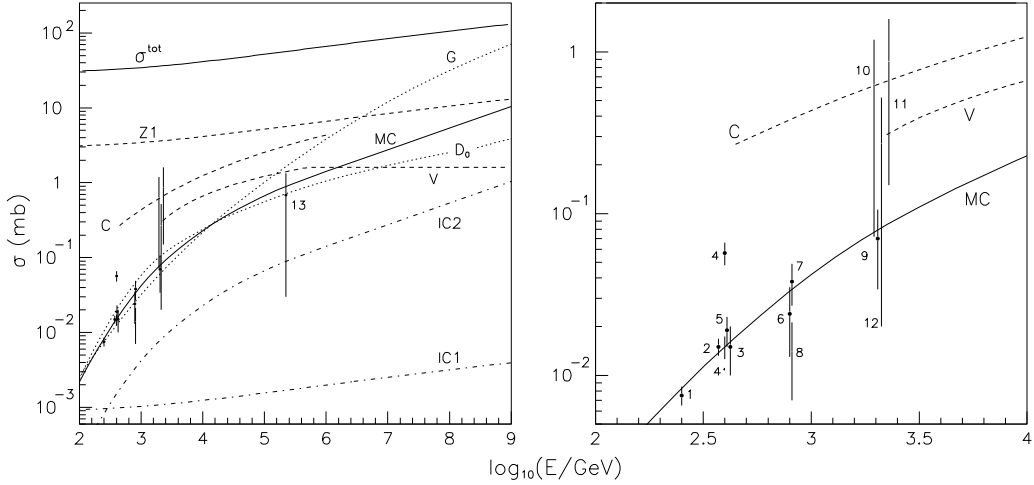}
\caption[junk]{\it Energy dependence of charm production cross section in $pp
(p\bar{p})$. The experimental data points are 1:\cite{E769a} 2:\cite{lebc400},
3:\cite{Jonker}, 4,4':\cite{Duffy}, 5:\cite{Fritze},  6:\cite{lebc800},
7:\cite{E653}, 8:\cite{E789} 9:\cite{Clark}, 10:\cite{r608},
11:\cite{Basile1,Basile3},  12:\cite{Basile2,Basile3} and 13:\cite{ua2}.  The
solid line is the result of our model. The dotted lines result from a naive
application of the $MRS$ parametrisations $G$ and $D_0$  of parton densities.
The dashed lines represent earlier models,  C from \cite{Castagnoli84}, V from
\cite{Volkova87} and Z1 from \cite{Zas93} which is 10\% of the total $pp$ cross
section $\sigma^{tot}$. The curves IC1 and IC2 are based on the intrinsic charm 
hypothesis (section~\ref{sec:IC}). 
Detailed discussions are in sections \ref{sec:charm} and \ref{sec:Comparison}.}
\label{fig:cross} 
\end{center} 
\end{figure}

The charm production cross section depends on the input for the parton  density
functions $f_i(x_i,Q^2)$ in Eq.\,(\ref{SIGMACC}).  With charm production being 
dominantly close to threshold  $\hat{s}=x_1x_2s > 4m_c^2 c^4$, the typical 
initial momentum fractions $x_i$ will decrease with increasing collision 
energy $s$. This is demonstrated in Fig.\,\ref{fig:5}, which shows the 
distribution of initial parton momentum fractions in charm production at 
different energies. At the highest energies, the parton densities are probed 
down to $x\sim 10^{-5}$ or even below.  The recent data from the $ep$ collider
HERA \cite{hera1,hera2} show a  significant increase at small $x$, $xf(x)\sim
x^{-a}$ and constrain the  parton densities down to $x\sim 10^{-4}$.  These
data, together with other data from previous deep inelastic scattering 
experiment as well as other processes, have been used in the  parametrization
$MRS\,G$ \cite{mrsg} of parton densities. The resulting  small-$x$ behaviour is
given by the power $a=0.07$ for sea quarks and $a=0.30$  for gluons. Since
$MRS\,G$ is the most recent parametrisation, using  essentially all relevant
experimental data, we use this as our standard choice.  To investigate
\cite{thunman} the dependence on the choice of parton density parametrisations,
we also applied the $MRS\,D_0$ \cite{mrsd0} with the small-$x$ behavior 
$x\,f(x)\sim const$, which before the HERA data  was an acceptable
parametrisation.  The effect on the total charm production cross section from
the choice of parton density parametrisation is illustrated in
Fig.\,\ref{fig:cross} with  curves resulting from these parametrisations. At
high energy there is a large dependence on the choice of parton  density
functions.  The difference between the $G$ and the $D_0$ parametrisations 
should however not be taken as a theoretical uncertainty. First of all the
$D_0$ parametrisation is known to be significantly below the small-$x$ HERA
data and gives therefore a significant underestimate  at large energies.
Secondly, the naive extrapolation of the $G$ parametrisation below the 
measured region $x\gsim 10^{-4}$ at rather small $Q^2$ ($\sim m_c^2$)  leads to
an overestimate. A flatter dependence like $x^{-\epsilon}$ with $\epsilon
\simeq 0.08$ as $x\,\to\,0$ can be motivated (\cite{SS} and references therein)
based on a connection to the high energy behaviour of cross sections in the
Regge framework. The implementation of this approach in {\sc Pythia} makes a
smooth transition  to this dependence such that the parton densities are
substantially  lowered for $x\lsim 10^{-4}$ leading to a substantial reduction
of the  charm cross section at large energies, as given by the solid curve in
Fig.\,\ref{fig:cross}.

Using this procedure we get a quite decent agreement between experimental charm
production data and the {\sc Pythia} simulation results  over a wide range of
energies (Fig.~\ref{fig:cross}).  A few comments on the data in 
Fig.\,\ref{fig:cross} are here in order.  A given experiment is only sensitive
to some channels and a limited  kinematical region. The total charm cross
section is therefore obtained by a rescaling with charm decay branching ratios
and by using assumed  shapes of the $x_F$ distributions to extrapolate to
unmeasured regions.  In particular, corrections to points 1,2,6 and 7 are
small  while they are large for point 9 and 13.  The bands 8,10,11 and 12
illustrates the uncertainty in these  experiments due to this extrapolation. 
In band 8 the uncertainty includes a scaling for including $D^{\pm}$-mesons
(taken from \cite{lebc800, E789}). Data-band 11 is based on $D^+\bar{D}$
identification, 12 on $\Lambda_c^+\bar{D}$ and 10 on $\Lambda_c^+$ 
identification. Furthermore, points 3, 4 and 5 are from beam dump experiments
on heavy nuclear targets without direct charm identification and have an
additional  uncertainty from the scaling with nuclear number. In point 4 the
scaling $A^{0.75}$ has been assumed, which we have rescaled in 4' to a
$A^1$-dependence in order to be consistent with the other beam dump 
experiments and with our model.  Data points 2 and 6 come from $pp$
interactions with explicit charm particle identification.   Although these
issues leave some uncertainty for each individual result,  the combination of
all data should give  a trustworthy knowledge on the charm cross section and
its energy dependence.

As indicated, there is an uncertainty related to the dependence of the charm
production cross section on the nuclear mass $A$.  Since the discussed pQCD
charm production involves hard  scattering processes with a small cross
section, one may argue that it does not only take place with nucleons at the
nuclear surface but also with nucleons in the interior. Consequently,  the
power $\alpha_c$  in $\sigma(pA\to c\bar{c}) = A^{\alpha_c} \sigma(pN\to
c\bar{c})$ may be larger than 2/3 and rather be 1, corresponding to
interactions in the whole nucleus. In fact, experiments give values of
$\alpha_c$  that are mainly close to 1 (see {\it e.g.} \cite{Vogt92} and
references therein), which is the value we have adopted in our calculations.
This is also in agreement with a number of recent experiments with both pion
and proton beams \cite{E789,E769b,WA82}. Using $\alpha_c=2/3$ instead would
lead to a reduction of the normalisation  for the prompt muon and neutrino
fluxes by the factor  $A^{2/3}/A=14.5^{-1/3}\,\approx\,0.4$. This is at most an
upper bound for the uncertainty of the $A$-dependence.

\subsection{Semileptonic decays of charmed hadrons}

The branching ratios for semileptonic charm decays used are $BR(D^{\pm} \to
e/\mu)=17\%$, $BR(D^0 \\ \to e/\mu)=8\%$,   $BR(D_s^{\pm} \to e/\mu)=8\%$,
$BR(\Lambda_c^+ \to e/\mu)=4.5\%$. These values agree with experimental
measurements as compiled by the  `Particle Data Group' \cite{pdb}.  To
distribute momenta in the charm hadron decay $H \to \ell \nu_{\ell} h$,  the
Monte Carlo \cite{pythia} uses the weak $V-A$ matrix element
\begin{equation}
\label{eq:weakmatrelem}
|{\cal M}|^2 = (p_Hp_{\ell})(p_{\nu}p_h)
\end{equation}
in terms of the involved four-vectors and neglecting decay product masses.  $H$
is the charm hadron and $h$ is the final hadron in a three-body decay and
generalized to represent the final state hadron system in case of more final
hadrons. The multiplicity and momenta within the $h$-system is
phenomenologically modelled \cite{pythia} and tested against data.  These
details, however, concern the final hadrons and are not important for our
purposes, because we only use the final leptons, which should be well accounted
for through the weak decay matrix element~(\ref{eq:weakmatrelem}) and the known
branching ratios.


\section{Possible non-perturbative origin of charm}
\label{sec:IC}

Although most of the charm production data from accelerator experiments can be
reasonably well understood from pQCD calculations, the uncertainty in the data
and the calculations cannot exclude some smaller non-perturbative contribution.
Charm production in pQCD is theoretically well defined and only has some
limited numerical uncertainty due to parameter values and NLO corrections
which, however, can be examined and controlled as discussed above. In contrast,
non-perturbative charm production is not theoretically well defined due to the
general problems of non-perturbative QCD. In particular, the absolute
normalisation has to be taken from comparison with data. Models for
non-perturbative charm production exist and some have been used in the context
of atmospheric muon and neutrino fluxes, as discussed in
section~\ref{sec:Comparison}. Here, we will investigate the consequences of the
hypothesis of intrinsic charm quarks in the nucleon wave function
\cite{Brodsky80}. Although being far from established, this idea has
theoretical motivation \cite{Brodsky80,Brodsky92} and some experimental data
can be interpreted as giving evidence
[62--67].  
It is therefore a serious model for a possible additional mechanism for charm
production with a non-perturbative origin. The results presented in this
section are based on a more general study \cite{IC-study} of intrinsic charm in
high energy collisions, to which we refer for more details of our
implementation in an explicit model.

The hypothesis of intrinsic charm (IC) amounts to assuming the existence of  a
\ccbar -pair as a non-perturbative component in the bound state  nucleon
\cite{Brodsky80}. This means that the Fock-state decomposition of,  {\it e.g.},
the proton wave function, $|p\rangle = A |uud\rangle + B |uudc\bar{c}\rangle +
...$, contains a small, but finite, probability $B^2$ for such  an intrinsic
quark-antiquark pair. This should be viewed as a quantum fluctuation of the
proton state.  The normalization of the heavy quark Fock component is the key
unknown, although it should decrease as $1/m_Q^2$.  Originally, a 1\% \
probability for charm was assumed, but later  investigations, \eg 
\cite{HoffmanMoore,Harris}, indicate a smaller but non-vanishing level.

Viewed in an infinite momentum frame, all non-perturbative and thereby 
long-lived components must move with essentially the same velocity in 
order that the proton can `stay together' for an appreciable time. 
The larger mass of the charmed quarks then implies that they take 
a larger fraction of the proton momentum. This can be quantified by
applying old-fashioned perturbation theory to obtain the momentum 
distribution \cite{Brodsky80} 
\begin{equation} 
P(p\to uudc\bar{c}) \propto 
\left[ m_p^2 - \sum_{i=1}^5 \frac{m^2_{\perp i}}{x_i} \right]^{-2}
\label{eq:IC-x}
\end{equation}
in terms of the fractional momenta $x_i$ of the five partons $i$ 
in the $uudc\bar{c}$ state.  
Neglecting the transverse masses of the light quarks in comparison to the charm
quark mass results in the momentum distribution 
\begin{equation}
P(x_1,x_2,x_3,x_c,x_{\overline{c}}) \propto 
\frac{x_c^2 x_{\overline{c}}^2}{(x_c + x_{\overline{c}})^2} 
\delta(1-x_1-x_2-x_3-x_c-x_{\overline{c}})
\label{eq:IC-simp}
\end{equation}
which favour large charm quark momenta as anticipated. In fact,
one obtains $\langle x_c \rangle =2/7$ by integrating out the light 
quark degrees of freedom $x_i$. 

A proton with such an intrinsic \ccbar \ quantum fluctuation can then  interact
with another hadron such that charmed particles are realised.  A hard
interaction with such a charm quark is certainly possible, but  the cross
section is then suppressed both by the small probability  of the fluctuation
itself and by the smallness of the  perturbative  QCD interaction. The charm
quarks may, however, also be put on shell  through non-perturbative
interactions that are not strongly suppressed \cite{Brodsky92}. This may lead
to a rate that is large enough to be of potential interest.   To estimate these
non-perturbative interactions we have constructed a  model \cite{IC-study}
based on refs.~\cite{Brodsky80,Vogt91,Vogt92}.

The formation of charm hadrons can occur through the following mechanisms. The
charm (anti)quark can hadronise into a $D$-meson as described by a normal
hadronisation function, similar to those successfully used in $e^+e^- \to
c\bar{c}$. Alternatively, the (anti)charm quark can coalesce with another quark
or diquark  from the $|uudc\bar{c}\rangle$ state to form a hadron.  Following
\cite{Vogt91,Vogt92} we use the recombination probabilities 50\% \ to form a
$\overline{D}$-meson and 30\% \ for a $\Lambda_c$, in which cases the remaining
$c$ or  $\bar{c}$ quark is assumed to hadronise separately from the proton
remnant. The probability to directly form a $J/\psi$ (\ie the \ccbar \ pair is
combined) is taken to be 1\%. The momentum of the hadron formed through
coalescence  is taken as the sum of the corresponding $x_i$'s, \eg  
$x_{\Lambda_c}=x_c+x_u+x_d$. The momentum distribution is then obtained by
folding Eq.\,(\ref{eq:IC-simp}) with the proper $\delta$ function, \eg 
$\delta(x_{\Lambda_c}-x_c-x_u-x_d)$, and integrating out all extra degrees of
freedom. The $c$ or $\bar{c}$ quarks  that does not coalesce with spectator
partons are hadronised  to $D$-mesons with a normal hadronisation function.
Since such a function is quite hard, we here approximate it with a
$\delta$-function to let the charm hadron take the whole charm
quark momentum given by Eq.\,(\ref{eq:IC-simp}), which is consistent with
low-$p_t$ charmed hadroproduction data  \cite{Vogt95}.

The shapes of the $x_F$-distributions for the charmed particles are thereby 
given (see \cite{IC-study}). They are quite hard, in fact harder than those for
charm from pQCD,  and therefore have the potential to contribute effectively at
high energies.  As mentioned, the main uncertainty in the intrinsic charm model
is the  absolute normalization of the cross section and its energy dependence. 
The magnitude of the cross section has been estimated \cite{Vogt92}  from data
at relatively low energies ($E_p= 200-400\,GeV$).  Since the process is
basically a soft non-perturbative process it may be  reasonable to assume that
its energy dependence is the same as that for  normal inelastic scattering
\cite{Brodsky92,IC-study}.  We therefore take as our first case
\begin{equation}
{\rm IC1:}\;\;\; \sigma_{IC}(s) =  3\cdot 10^{-5} \sigma_{inel}(s)
\label{eq:IC1}
\end{equation}
with normalisation from \cite{Vogt92} and 
shown as curve IC1 in Fig.\,\ref{fig:cross}.  
Alternatively, one might argue that there is a stronger energy dependence
related to some threshold behavior for putting the charm quarks on their 
mass shell. We make a very crude model for this by taking the intrinsic charm
cross section to be a constant fraction of the pQCD charm cross section
\begin{equation}
{\rm IC2:}\;\;\; \sigma_{IC}(s) =  0.1\: \sigma_{pQCD}(s)
\label{eq:IC2}
\end{equation}

as shown by curve IC2 in Fig.\,\ref{fig:cross}. This is similar to the low
energy ($200-800\,GeV$) treatement in~\cite{Vogt91}. The normalisation is here
fixed to be the same as IC1 at the low energy where evidence is claimed for
intrinsic charm \cite{Vogt92}. There is, however, some indication against such
an increased cross section, as in IC2, since no evidence for $J/\psi$ from
intrinsic charm was found in an  experiment \cite{noIC} at a somewhat higher
energy ($800\,GeV$ proton beam).

The dependence on nuclear mass number should (in both cases) essentially be 
$A^{2/3}$ reflecting the soft nature of the hadron-hadron interaction.  Note,
that the intrinsic charm quarks can be released through interactions with the
other parts of the $|uudc\bar{c}\rangle$ state, as demonstrated  clearly in
case of $J/\psi$ production after the remaining proton state  has been
`stripped off' in the interaction \cite{Brodsky92}.

\begin{figure}
\begin{center}
\includegraphics[width=13cm]{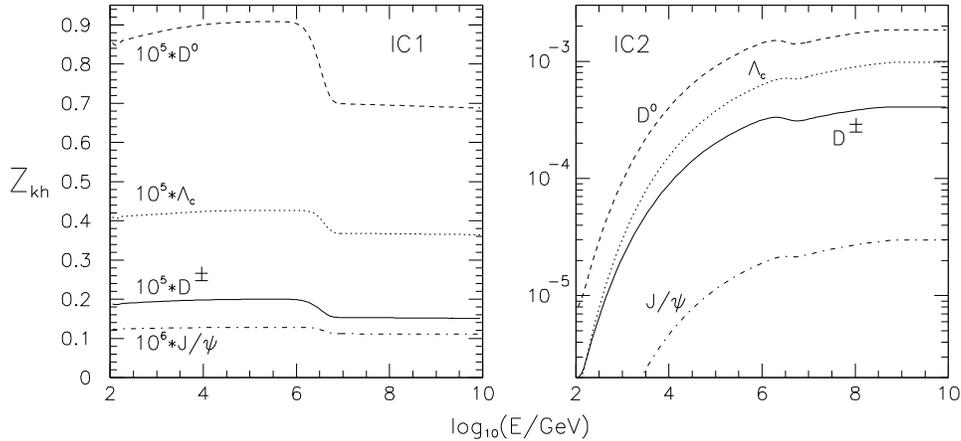}
\caption[junk]{\it Energy-dependence of 
production $Z_{kh}$-moments, Eq.\,(\protect\ref{eq:zmom}), 
for incoming particle $k=proton$ producing charmed hadron
$h=D^0,D^{\pm},\Lambda_c$ and $J/\psi$ through the 
intrinsic charm mechanism with the two assumed energy dependences 
IC1 and IC2.}
\label{fig:zmomic} 
\end{center}
\end{figure}

The intrinsic charm model provides simple and scaling $x_F$-spectra for the 
charmed particles which makes the analytic method suitable for calculating the
fluxes of muons and neutrinos from their decays. The charm production
$Z$-moments are calculated according to Eq.\,(\ref{eq:zmom}) by numerical
integration and shown in Fig.\,\ref{fig:zmomic}.  The mild energy dependence of
the IC1 case gives essentially constant  $Z$-moments, except for a marked step
due to the change of the slope $\gamma$  in the primary spectrum. The step is
more pronounced here compared to charm  from pQCD (see Fig.\,\ref{fig:zmom})
where the $x_F$-distribution is both  energy-dependent and extending to smaller
values.  The stronger energy dependence in IC2 is reflected in the
corresponding  $Z$-moments, where the strong variation at energies below
$\sim10^5\,GeV$ may  cast some doubt on the reliability of the analytic method
in this region.

\begin{figure}[thb]
\begin{center}
\includegraphics[width=12cm]{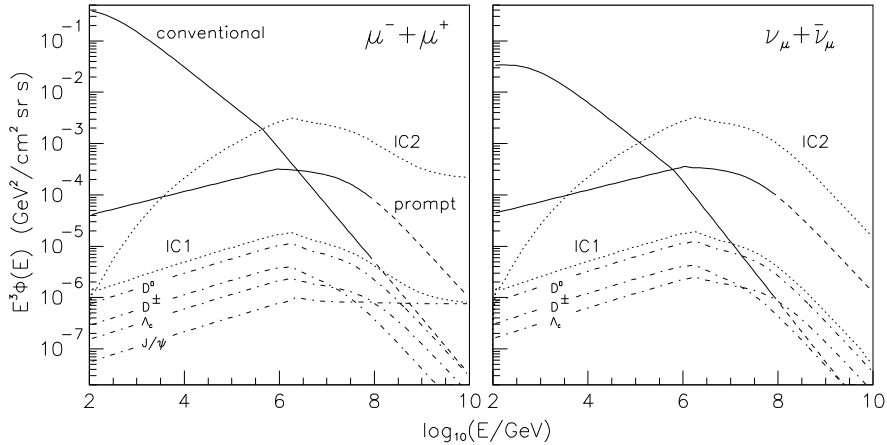}
\caption[junk]{\it 
Fluxes (cf. fig.\,3) of muons and neutrinos from the intrinsic charm model
with the two assumed energy dependences IC1 and IC2;
for IC1 the contributions from different charm particles are given.
Shown for comparison are our standard results (full curves) for 
conventional and prompt fluxes as given by the parametrisation 
Eq.\,(\ref{eq:fiteq}) and extrapolated to higher energies (dashed).}
\label{fig:ic} 
\end{center}
\end{figure}

The lepton fluxes are then obtained by using Eq.\,(\ref{eq:interpol}) and the
regeneration and decay $Z$-moments calculated in section\,4. The results are
displayed in Fig.\,\ref{fig:ic} and compared to those from our pQCD
calculation. The milder, and more conservative, energy dependence (IC1) of the
intrinsic charm cross section gives a result which is only a small ($\sim
10\%$) correction to the pQCD result, except at super-high energies where the
rate is extremely small and not measurable in a foreseeable future. Note that
the  $J/\psi$ contribution is here becoming important, since the $J/\psi$ flux
is not attenuated through interactions due to the high critical energy
(Table\,\ref{tab:epsilon}).  This raises the question of how well the
normalisation of the otherwise small $J/\psi$ contribution is known. With the
strong energy  dependence assumed in IC2, the intrinsic charm result exceeds
the  pQCD one already at lepton energies around $10^4\: GeV$. Although the
energy dependence of this IC2-model is, as mentioned, rather {\em ad hoc} and
may be disfavoured by data,  it illustrates the large theoretical uncertainty
associated with the intrinsic charm model when extrapolated to the high 
energies of cosmic ray interactions.


\section{Comparison with previous model calculations}
\label{sec:Comparison}
In sections 3 and 4 we have obtained atmospheric muon and neutrino fluxes with
two different methods: via a Monte Carlo simulation of the hadronic cascade and
via approximate analytical expressions with energy-dependent $Z$-moments. We
were satisfied that the two methods gave consistent results. Here we want to
compare our results with those obtained with different models for particle
interactions in the atmosphere. In particular, we focus on the prompt muon and
neutrino fluxes arising from different charm production mechanisms. 

\begin{figure}[tbh]
\begin{center}
\includegraphics[width=16cm]{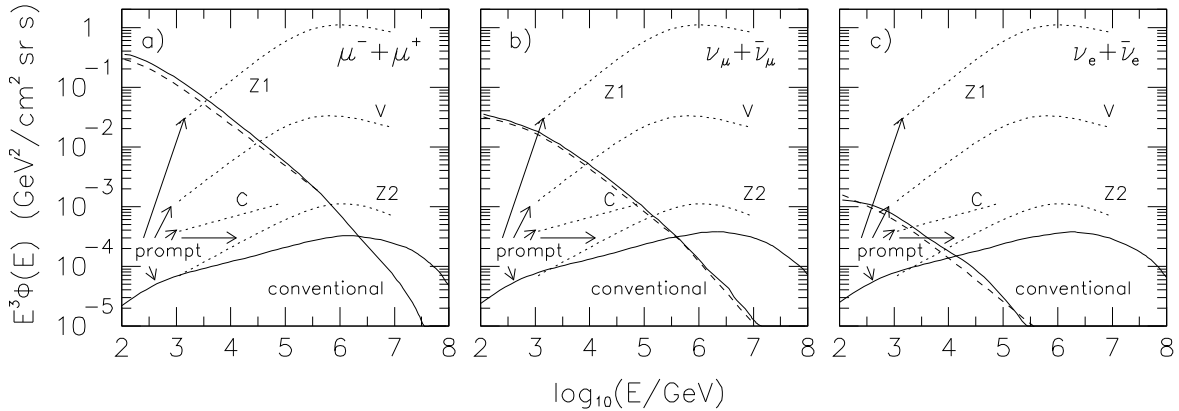}
\caption[junk]{{\it Prompt and conventional muon and neutrino fluxes from 
our cascade 
simulation (solid curves) compared with earlier model calculations as discussed
in the text.}}
\label{fig:8} 
\end{center}
\end{figure}

\begin{figure}
\begin{center} 
\includegraphics[width=14cm]{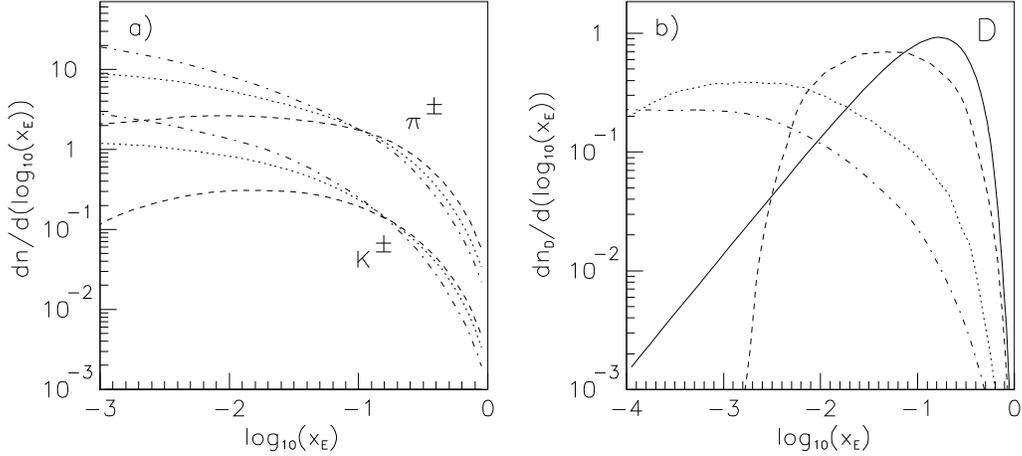}
\caption[junk]{{\it Distribution in fractional energy $x_E$ of produced 
mesons: (a) $\pi^\pm$ and $K^\pm$ mesons, (b) $D$ mesons. 
Dashed, dotted and dashed-dotted lines are obtained from
simulations with {\sc Pythia} at proton beam energies of 
$10^3$\,GeV, $10^6$\,GeV and $10^9$\,GeV, respectively. 
The solid line represents Feynman scaling using 
Eq.\,(\protect\ref{eq:qgsm}) with $\alpha_D=5$. 
($D$-meson curves are normalized to unit area.)}}
\label{fig:6}
\end{center}
\end{figure}

Earlier calculations of the conventional muon and neutrino fluxes 
\cite{Gaisser90,Volkova80,GSB88,Lipari93} agree well with our results as shown
in Fig.\,\ref{fig:8}. The conventional muon flux  from Gaisser \cite{Gaisser90}
is shown as a dashed line in Fig.\,\ref{fig:8}a as far in energy as it is
applicable.  Also shown as dashed lines in Fig.\,\ref{fig:8}bc are the
conventional neutrino fluxes  from Volkova \cite{Volkova80}. In contrast to
these models, ours does not obey Feynman scaling.  The scaling violations are
apparent in Fig.\,\ref{fig:6}a and from the deviation from a constant value of
the production  $Z$-moments (Fig.\,\ref{fig:zmom}). Nevertheless, they are
small enough that  when folding  everything together (initial spectrum, cascade
interactions and decays)  the resulting conventional muon and neutrino fluxes
agree well with those  models. We have thus confirmed previous results  by an
independent calculation based on a new approach using Monte Carlo  simulations
to more fully take into account the atmospheric cascade  interactions producing
secondary particles decaying into muons and neutrinos.

Concerning previous estimates of the flux of prompt muons and neutrinos, there
are variations between different model calculations of up to a few  orders of
magnitude, as illustrated in Fig.\,\ref{fig:8}. One should note that prompt
fluxes are direction  independent up to the charm particle critical energy
$\sim10^7$\,GeV  (Table\,\ref{tab:epsilon}) and therefore directly comparable
independently of whether the horizontal or vertical direction has been
considered in these estimates.  Furthermore, due to the charmed particle decay
kinematics and the same  branching ratios for the semi-leptonic decays into
electrons and muons, the prompt muon and neutrino fluxes are essentially the
same (cf.~the decay $Z$-factors in Table~\ref{tab:zdec}). Therefore, the curves
for prompt muons in Fig.\,\ref{fig:8}a are also taken to  represent the prompt
neutrinos in Fig.\,\ref{fig:8}bc (except for our own curves, which are
calculated separately). The comparison in Fig.\,\ref{fig:8} shows that previous
results are in general substantially larger than our results based on pQCD.
Even the highest flux from our extreme version IC2 of the intrinsic charm 
hypothesis (Fig.\,\ref{fig:ic}) is lower than most previous calculations. These
large differences are due to different models for charm production, both
regarding the magnitude and the energy dependence of the cross section and the
distribution in longitudinal momentum fraction $x_F$ of the charmed particles.

The curves labeled V in Fig.\,\ref{fig:8} are from the calculation  by Volkova
\etal\ \cite{Volkova87},  applying the so-called `quark-gluon string model'
(QGSM) \cite{qgsm}  (not to be confused with the Lund string model
\cite{lund}).  It uses a parametrised energy dependence of the charm cross
section,  curve labeled V in Fig.\,\ref{fig:cross}, normalized to early
experimental data which  are substantially above more recent measurements.  The
charm particle energy spectrum is assumed to obey Feynman scaling  and has the
form  \begin{equation} \label{eq:qgsm} dN/dx_F\sim (1-x_F)^{\alpha}
\end{equation} with $\alpha_D=5$ and $\alpha_{\Lambda_c}=0.4$ for  $D$-mesons
and $\Lambda_c$-baryons, respectively.

The curves marked Z1 in Fig.\,\ref{fig:8} are from Zas \etal\,\cite{Zas93} and
illustrates an extreme model where the charm cross section is simply taken as
10\% \ of the total inelastic cross section (cf.~Fig.\,\ref{fig:cross}). This
is substantially higher  than all charm data as shown in Fig.\,\ref{fig:cross}.
This model uses the scaling $x_F$-distribution of Eq.\,(\ref{eq:qgsm}) with
$\alpha_D=3$ and  $\alpha_{\Lambda_c}=1$. Castagnoli \etal\,\cite{Castagnoli84}
obtained the result marked C in Fig.\,\ref{fig:8} using a parametrised energy
dependent charm cross section shown by curve C in Fig.\,\ref{fig:cross} based
on some early data (band marked 11) that are higher than  later measurements.
Again, the differential spectra are of the form  Eq.\,(\ref{eq:qgsm}) using
$\alpha_D=5$ and $\alpha_{\Lambda_c}=0.4$. The curves marked Z2, from Zas
\etal\, \cite{Zas93}, correspond to charm  {\em quark} production calculated
with leading order pQCD matrix elements  using relatively hard parton
distributions. This spectrum would be  softened by taking hadronisation into
account and thereby become even  closer to our result, as expected since they
are based on the same  pQCD processes.

The first important difference between our model and previous ones lies in the 
magnitude and energy dependence of the charm production cross section.  As
demonstrated in Fig.\,\ref{fig:cross} our model reproduces available
data on charm production cross sections, but the other models do not. 
In some cases one may have been mislead in the construction of the models 
by the early charm measurements that turned out to be substantially higher 
than the measurements done later. 

Another important reason for our lower flux is the strong breaking of Feynman
scaling as demonstrated in Fig.\,\ref{fig:6}b.  The $x_F$ distributions for
$D$-mesons produced at three different energies in our model are here compared
with the energy-independent distribution in Eq.\,(\ref{eq:qgsm}) with
$\alpha_D=5$. The Feynman scale breaking in our model arises in the
perturbative charm quark production, but is also influenced by the
hadronization model.  As discussed in section 5.2, charm quark production is
dominant close to  threshold, $\hat{s}=x_1x_2s > 4m_c^2 c^4$. This effect does
not disappear with  increasing energy, but is rather enhanced with parton
densities that increase  at small $x$.  This leads to a scale breaking with the
charm quark $x_F$ distribution in the  symmetric nucleon-nucleon cms becoming
softer around $x_F=0$  with increasing cms energy.  In the Lund hadronization
model, the charm quark is connected by a colour  string to a spectator parton.
In the hadronization of this string, the produced charm hadron may obtain a
larger longitudinal momentum than the charm quark,  due to the momentum
contribution from the parton  with which it is joined. The string may even have
so small invariant mass that it directly produces a charmed hadron, \ie the
charm quark effectively  coalesces with a spectator parton into a charmed
hadron. This latter process naturally happens particularly at small overall cms
energies. Since these forward-`pulling' hadronization effects become less
important with increasing  collision energy, they also contribute to the
Feynman scale breaking.

\begin{figure}[b]
\begin{center} 
\includegraphics[width=7cm]{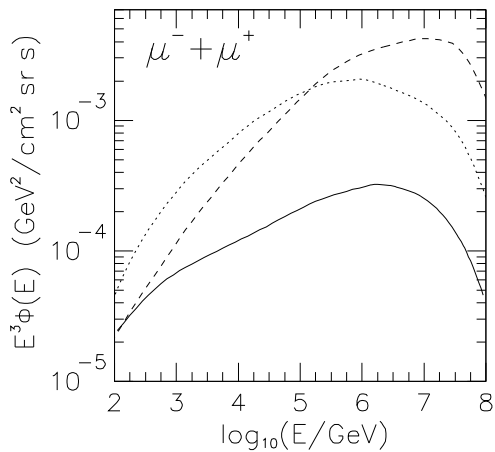}
\caption[junk]{{\it The prompt muon flux in our model (solid line) and after 
re-distributing the generated charm particles to obey Feynman scaling 
using Eq.\,(\protect\ref{eq:qgsm}) (dashed line) and then also renormalizing
the cross 
section to that of curve V in Fig.\,\ref{fig:cross} from \cite{Volkova87} 
(dotted curve).}}
\label{fig:11} 
\end{center}
\end{figure}

The effect on the prompt muon flux from this Feynman scale breaking is shown
in  Fig.\,\ref{fig:11}. Our normal result is here compared with the results
from a  modification of our model, where in each event the charmed particles 
$D$ and $\Lambda_c$ are redistributed according to the scaling distribution in 
Eq.\,(\ref{eq:qgsm}) with $\alpha_D=5$ and $\alpha_{\Lambda_c}=0.4$.  Clearly,
the Feynman scale breaking softens the spectrum considerably.  (In comparing
these curves one should note that there is no conserved  integral of the
$E^3$-weighted flux.) To examine the effect of the energy dependence of the
overall charm cross section, we have, in addition to using this scaling
$x_F$-distribution, renormalized our simulated charm events to mimic the cross
section in the model by Volkova \etal~\cite{Volkova87} mentioned above and
shown by curve V in  Fig.\,\ref{fig:cross}. Since this cross section is larger
than in our model at energies below  $\sim 10^6$\,GeV and smaller above, this
change flattens the muon spectrum in Fig.\,\ref{fig:11}. With these two changes
in the spirit of ref.~\cite{Volkova87}, we obtain the same shape of the prompt
muon flux as in \cite{Volkova87} but with a lower overall normalization. The
resulting spectra are, however, in reasonable agreement with the  calculation
of Castagnoli \etal~\cite{Castagnoli84} using a similar  approach as
ref.\,\cite{Volkova87}.  A more recent calculation based on the QGSM by
Gonzalez-Garcia \etal~\cite{Gonzalez} gives fluxes that are comparable to the
fluxes predicted by Castagnoli \etal~\cite{Castagnoli84}.

The calculation by Bugaev \etal~\cite{Bugaev} resulted in overall  prompt
fluxes slightly lower than in ref.\,\cite{Volkova87}.  They considered Feynman
scaling violations in charm production through a  phenomenological equation and
obtained higher fluxes in the non-scaling  case than in the scaling case, \ie
opposite to the effect we find and  have just described. However, their way of
introducing the energy dependence in the $x_F$  distribution does not preserve
the overall normalization, \ie the integral  of the $x_F$-distribution. This
means that, in comparison with our model,  there is not the same clear
separation between the overall charm cross section  normalization and the charm
particle $x_F$ distribution.

Since most of the earlier calculations are based on non-perturbative charm
production mechanisms a comparison with our intrinsic charm model in section 
\ref{sec:IC} is of interest, \ie comparing Figs.\,\ref{fig:ic} and \ref{fig:8}.
Intrinsic charm should be considered as a process in addition to the standard
pQCD one, but the sum of their resulting fluxes is still lower than,  {\it
e.g.}, Volkova \etal\ \cite{Volkova87}.  To get a similarly high flux the rate
of intrinsic charm must be increased  substantially, about a factor 1000 for
IC1 and 10 for IC2.  Such rates are incompatible with the experimental limits
on intrinsic charm  and with measured inclusive charm cross sections
(Fig.\,\ref{fig:cross}) and $x_F$-distributions.

This discussion has demonstrated significant effects on the high energy  prompt
muons and neutrinos depending on the assumptions made in the charm production
model employed. The models in ref.~\cite{Castagnoli84,Volkova87}  give cross
sections above more recent charm production data  (Fig.\,\ref{fig:cross})  and
apply simple Feynman scaling $x_F$-distributions.   The model used by us, on
the other hand, gives a fair description of measured  charm production cross
sections (Fig.\,\ref{fig:cross}) and applies  well-motivated  charm particle
momentum distributions with significant Feynman scaling  violations.


\section{Conclusions and outlook}

We have studied the production of neutrinos and muons  in the atmosphere by
collisions of cosmic rays with air nuclei, paying special attention to  muons
and neutrinos coming from decays of charmed particles (prompt fluxes). Two
methods have been used to calculate the fluxes: a Monte Carlo simulation of the
hadronic cascade in the atmosphere and an interpolation of asymptotic solutions
to the transport equations. In both methods the {\sc Pythia} Monte Carlo
program has been used to simulate the primary collision and following cascade
interactions. Results for the two methods are consistent. They agree with
previous calculations of the conventional fluxes from decays of pions and
kaons, but give substantially lower prompt components. This is due to 
different models for charm production, both regarding the energy dependence of
the cross section and the longitudinal momentum distribution of the charmed
particles. Whereas previous models give charm production cross sections above
collected recent data and apply Feynman  scaling for the longitudinal momentum
distributions, our model gives a fair  description of measured charm production
cross sections and applies well-motivated charm particle momentum distributions
with significant Feynman scaling violations. There is still some uncertainty
(as discussed in section 5.2) when extrapolating this charm cross section
calculation to the very high energies ($10^9$\,GeV) needed for this study.
Here, one cannot at present exclude a non-negligible contribution from some 
unconventional non-perturbative production mechanism. We investigate one such
mechanism,  namely that of intrinsic charm which has some theoretical
motivation and  some indications from data. This mechanism is likely to give a
very small contribution to the total charm cross section, but the poorly known 
normalisation and energy dependence prevents a reliable prediction.  Although
disfavoured, a contribution $\sim 10\%$ of the pQCD charm cross  section is
presently not excluded, which through the harder momentum spectrum  would lead
to a dominant contribution of leptons at very high energies.

We find that prompt muons and muon-neutrinos overcome the conventional fluxes 
at an energy of $10^6$\,GeV, which is substantially higher than in some
earlier  estimates. According to our results, it will therefore be harder to
use measurements of the prompt  atmospheric fluxes to estimate the total charm
production cross section  at high energy. The situation is slightly different
in the case of the  electron-neutrinos, for which prompt fluxes dominate above
$10^{5}$\,GeV. The electron-neutrino flux is, however, experimentally more
difficult to  measure and it is therefore a challenge to obtain the data needed
to  derive the charm production cross section.

\begin{figure}
\begin{center}
\includegraphics[width=8cm]{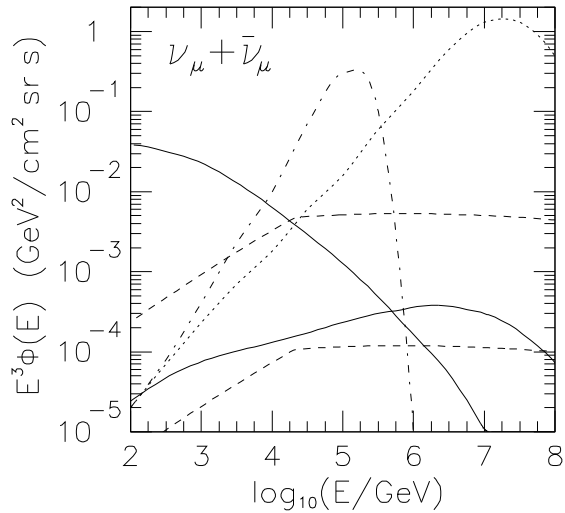}
\caption[junk]{{\it Vertical fluxes 
of conventional and prompt atmospheric muon-neutrinos
from our simulation (solid lines) compared to some astrophysical sources:
the flux from cosmic ray interactions with the interstellar medium as 
derived from \cite{Domokos94} (dashed upper curve: in the direction of the
galactic center; dashed lower curve: orthogonal to the galactic plane)
and the estimated diffuse fluxes from active galactic nuclei 
(dotted line \cite{Szabo92}, dash-dotted line \cite{Sikora92}).}}
\label{fig:9} 
\end{center}
\end{figure}

On the positive side, the lower atmospheric neutrino fluxes we predict are  a
less severe background to measurements of neutrinos from astrophysical 
sources  (for a review on these see ~\cite{Berezinski}). To illustrate this, we
show in  Fig.\,\ref{fig:9} the vertical fluxes of conventional and prompt
atmospheric muon-neutrinos calculated by us (solid lines) together with
expected neutrino fluxes from such sources. Cosmic ray interactions with the
interstellar medium produce neutrino  fluxes through processes similar to the
atmospheric case and we show results derived from \cite{Domokos94} in the
direction of the galactic center (dashed upper curve) and   orthogonal to the
galactic plane (dashed lower curve).  Two estimates of diffuse neutrino fluxes
from active galactic nuclei are also shown (dotted line from
ref.~\cite{Szabo92}  and dash-dotted line from ref.~\cite{Sikora92}).  At high
energies ($\gsim 10^5$\,GeV), all of these fluxes  are in excess of our
predicted atmospheric neutrino background. This provides interesting prospects
for large scale neutrino telescopes  to detect high energy neutrinos from
cosmic sources.


\section*{Acknowledgements}

We are grateful to T.~Sj\"{o}strand for useful discussions and to J.~Rathsman
for a critical reading of the manuscript. P.~Gondolo wishes to thank the
Uppsala and Stockholm Universities for support during his visits. This work has
been partially funded by the Swedish Natural Science Research Council and the
European Community through the Theoretical Astroparticle Network under contract
No. CHRX-CT93-0120 (D.G.~12~COMA).



\begin{thebibliography}{99}

\bibitem{Gaisser90} T. K. Gaisser, {\it Cosmic Rays and Particle Physics\/}
(Cambridge University Press, Cambridge, 1990).

\bibitem{amanda} {\sc AMANDA} Collaboration, in {\it 23rd International
Cosmic Ray Conference,} Calgary, Canada, July 1993, Vol.4, p.561.

\bibitem{baikal} G. V. Domogatsky, in {\it TAUP 93,} ed.\ C. Arpesella,
E. Bellotti and A. Bottino, Nucl.\ Phys.\ (Proc.\ Suppl.) {\bf B35} (1994) 290.

\bibitem{dumand} J. G. Learned, in {\it Neutrino 92,} ed. A Morales,
Nucl.\ Phys.\ (Proc.\ Suppl.) {\bf B31} (1993) 
456.

\bibitem{nestor} L. K. Resvanis, in {\it TAUP 93,} ed.\ C. Arpesella, E.
Bellotti and A. Bottino, Nucl.\ Phys.\ (Proc.\ Suppl.) {\bf B35} (1994) 294.

\bibitem{thunman} M. Thunman, Licentiate thesis, Uppsala University,
TSL/ISV-94-0097.

\bibitem{Volkova80} L. V. Volkova, Yad.\ Fiz.\ {\bf 31} (1980) 1510 
[Sov.\ J.\ Nucl.\ Phys.\ {\bf 31} (1980) 784].

\bibitem{GSB88} T. K. Gaisser, T. Stanev and G. Barr, Phys.\ Rev.\ {\bf D38}
(1988) 85; G. Barr, T. K. Gaisser and T. Stanev, Phys.\ Rev.\ {\bf D39} (1989)
3532.

\bibitem{Lipari93} P. Lipari, Astropart.\ Phys.\ {\bf 1} (1993) 195.

\bibitem{Bugaev-at-nestor} For a review of the 1993 status, see
E.V. Bugaev \etal, {\it Proceedings of the 3rd NESTOR International
Workshop}, Pylos (Greece), 1993, ed.\ L.K. Resvanis, p.~268.

\bibitem{Rhode} W. Rhode, {\it TAUP93}, Gran Sasso (Italy), 1993, ed.\
C. Arpesella, E. Bellotti and A. Bottino, Nucl.\ Phys.\ B (Proc.\
Suppl.) 35 (1994) 250.

\bibitem{Macro} M. Ambrosio \etal\ (MACRO Collaboration), Phys. Rev. {\bf D52}
(1995) 3793.

\bibitem{Lvd} M. Aglietta \etal\ (LVD Collaboration), Astropart.\ Phys.\ {\bf 3}
(1995) 311.

\bibitem{Volkova83} L.V. Volkova and G.~T.~Zatzepin, Yad.\ Fiz.\ {\bf 37} (1983)
353 [Sov.\ J.\ Nucl.\ Phys.\ {\bf 37} (1983) 212].

\bibitem{Castagnoli84} C.~Castagnoli \etal, Nuovo Cimento {\bf A82} (1984) 78.

\bibitem{Inazawa86} H.~Inazawa, K.~Kobayakawa and T.~Kitamura, Nuovo Cimento
{\bf C9} (1986) 382.

\bibitem{Volkova87} L.V. Volkova, W. Fulgione, P. Galeotti and O. Saavedra,
 Nuovo Cimento {\bf C10} (1987) 465.

\bibitem{Bugaev} E.V.~Bugaev, E.S.~Zaslavskaya, V.A.~Naumov and S.I.~Sinegovsky,
in {\it 20th Int.\ Cosmic Ray Conference,} Moscow 1987, Vol.~6, p.~305.

\bibitem{Bugaev89} E.V.~Bugaev, V.A.~Naumov, S.I.~Sinegovsky and
E.S.~Zaslavskaya,
Izv. Akad. Nauk SSSR, Ser. Fiz. {\bf 53} (1989) 342 [Bull. Acad. of Sci. of the
USSR, Phys. Ser. {\bf 53} (1989) 135].

\bibitem{Zas93} E.~Zas, F.~Halzen and R.A.~V\'{a}zques, Astropart.\
Phys.\ {\bf 1} (1993) 297.

\bibitem{Gonzalez} M.~C.~Gonzalez-Garcia, F.~Halzen, R.A.~V\'{a}zques and
E.~Zas, Phys. Rev. {\bf D49} (1994) 2310.

\bibitem{Pal92} P.~Pal and D.~P.~Bhattacharyya, Nuovo Cimento {\bf C15} (1992)
401.

\bibitem{JACEE} T.H.~Burnett \etal, {\it Proceedings of the XXI International
Cosmic Ray Conference, Adelaide, 1990}, Vol.~3, p.~101.

\bibitem{MACRO} S.~Ahlen \etal, Phys. Rev. {\bf D46} (1992) 895.

\bibitem{Honda} M.~Honda, T.~Kajita, K.~Kasahara and S.~Midorikawa, {\it
Calculation of the Flux of Atmospheric Neutrinos}, ICR-Report-336-95-2.

\bibitem{Gregory82} A. Gregory and R. W. Clay, in {\it CRC Handbook of
Chemistry and Physics\/} (CRC Press Inc., Boca Raton, Florida, 1982--1983) p.\
F-175.

\bibitem{Allen83}  C. W. Allen, {\it Astrophysical Quantities\/} (The Athlone
Press, London, 1983).

\bibitem{formlength}
J.~Ashman \etal, Z.~Physik {\bf C52} (1991) 1 and references therein.

\bibitem{helios89}  J.~Shukraft \etal, HELIOS Collaboration, Nucl.\ Phys.\
{\bf A498} (1989) 79c.

\bibitem{Hoang94} T. F. Hoang,
 Z.\ Phys.\ {\bf C61} (1994) 341.

\bibitem{pythia} T. Sj\"ostrand, {\sc PYTHIA}~5.7 \& \ {\sc JETSET}~7.4, 
Comput.\ Phys.\ Commun.\ {\bf 82} (1994) 74

\bibitem{lund} B.~Andersson, G.~Gustafson, G.~Ingelman and T.~Sj\"{o}strand,
Phys. Rep. {\bf 97} (1983) 33.

\bibitem{pdb} Particle Data Group, Phys. Rev. {\bf D50} (1994) 1.

\bibitem{Frazer72} W. R. Frazer \etal, Phys.\ Rev.\ {\bf D5} (1972)
1653.

\bibitem{Garraffo73} Z. Garraffo, A. Pignotti and G. Zgrablich, Nucl.\ 
Phys.\ {\bf B53} (1973) 419.

\bibitem{mrsg} A.~Martin, R.~Roberts and J.~Stirling, Phys. Lett. {\bf B354}
(1995) 155.

\bibitem{Nason1} P.~Nason, S.~Dawson and R.K.~Ellis, Nucl. Phys. {\bf B327}
(1989) 49; {\it ibid.} {\bf B335} (1990) 260.

\bibitem{Nason2} M.L.~Mangano, P.~Nason and G.~Ridolfi, Nucl. Phys. {\bf B373}
(1992) 295.

\bibitem{Cifarelli} L.~Cifarelli, E.~E\c{s}kut and Yu.M.~Shabelski, Nuovo 
Cimento
{\bf A106} (1993) 389.

\bibitem{hera1} H1 Collaboration: I.~Abt \etal, Nucl. Phys. {\bf B407} (1993)
515; T.~Ahmed \etal, .Nucl. Phys. {\bf B439} (1995) 471.

\bibitem{hera2} ZEUS Collaboration: M.~Derrick \etal, Phys. Lett. {\bf B316}
(1993) 412; Z.~Phys. {\bf C65} (1995) 379, Phys. Lett. {\bf B345} (1995) 576.

\bibitem{mrsd0} A.~Martin, R.~Roberts and J.~Stirling, Phys. Lett. {\bf B306}
(1993) 145.

\bibitem{SS} G.A.~Schuler and T.~Sj\"{o}strand, Nucl. Phys. {\bf B407} (1993)
539.

\bibitem{E769a} P. E.~Karchin, E769 Collaboration, {\it Current Issues in Open
Charm Hadroproduction and New Preliminary Results from Fermilab E769},
FERMILAB-Conf-95/053-E.

\bibitem{lebc400} M.~Aguilar-Benitez \etal, Phys. Lett. {\bf B189} (1987) 476.

\bibitem{Jonker} M.~Jonker \etal, Phys. Lett. {\bf 96B} (1980) 435.

\bibitem{Duffy} M.\,E.~Duffy \etal, Phys. Rev. Lett. {\bf 57} (1986) 1522.

\bibitem{Fritze} P.~Fritze \etal,  Phys. Lett. {\bf 96B} (1980) 427.

\bibitem{lebc800} R.~Ammar \etal, Phys. Rev. Lett. {\bf 61} (1988) 2185.

\bibitem{E653} K.~Kodama \etal, Phys. Lett. {\bf 263B} (1991) 573.

\bibitem{E789} M. J. Leitch \etal, Phys. Rev. Lett. {\bf 72} (1994) 2542.

\bibitem{Clark} A.\,G.~Clark \etal, Phys. Lett. {\bf 77B} (1978) 339.

\bibitem{r608} P.~Chauvat \etal, Phys. Lett. {\bf B199} (1987) 304.

\bibitem{Basile1} M.~Basile \etal, Nuovo Cimento {\bf 67A} (1982) 40.

\bibitem{Basile3} A.~Contin \etal, in {\it Int. Conf. on High Energy Physics,} 
Lisbon 1981, eds. J. Dias de Deus and J. Soffer, p.~835.

\bibitem{Basile2} G.~Bari \etal, Nuovo Cimento {\bf 104A} (1991) 571.

\bibitem{ua2} O.~Botner \etal, Phys. Lett. {\bf B236} (1990) 488.

\bibitem{E769b} A. Alves \etal, Phys. Rev. Lett. {\bf 70} (1993) 722.

\bibitem{WA82} M. Adamovich \etal, Phys. Lett. {\bf 284B} (1992) 453.

\bibitem{Brodsky80} S.J.~Brodsky, P.~Hoyer, C.~Peterson and N.~Sakai, Phys. 
Lett. {\bf B93} (1980) 451.\\
S.J.~Brodsky, C.~Peterson and N.~Sakai, Phys. Rev. {\bf D23} (1981) 2745.

\bibitem{Brodsky92} S.J.~Brodsky \etal, Nucl. Phys. {\bf B369} (1992) 519. 

\bibitem{Vogt91} R.~Vogt, S.J.~Brodsky and P.~Hoyer, Nucl. Phys. {\bf
B360} (1991) 67. 

\bibitem{Vogt92} R.~Vogt, S.J.~Brodsky and P.~Hoyer, Nucl. Phys. {\bf
B383} (1992) 643. 

\bibitem{HoffmanMoore} E.~Hoffman and R.~Moore, Z.~Phys. {\bf C20} (1983) 71.

\bibitem{Harris} B.W. Harris, J. Smith and R. Vogt, {\it Reanalysis of the EMC
Charm Production Data with Extrinsic and Intrinsic Charm at NLO},
FSU-HEP-951030, ITP-SB-95-15, LBL-37266.

\bibitem{Vogt2phi} R. Vogt and S.J. Brodsky, Phys. Lett. {\bf B349} (1995) 569.

\bibitem{Vogt95} R.~Vogt and S.J.~Brodsky, {\it Charmed Hadron Asymmetries in
the Intrinsic Charm Coalescence Model}, SLAC-PUB-95-7068, LBL-37666.

\bibitem{IC-study} G.~Ingelman and M.~Thunman, in preparation.

\bibitem{noIC} M.~S.~Kowitt \etal, Phys. Rev. Lett. {\bf 72} (1994) 1318.

\bibitem{qgsm} A.B.~Kaidalov and O.I.~Piskunova, Z.~Physik {\bf C30} (1986) 145

\bibitem{Berezinski} V.S.~Berezinsky, S.V.~Bulanov, V.A.~Dogiel and
V.S.~Ptuskin, {\it Astrophysics of Cosmic Rays}
(North Holland, Amsterdam, 1990).

\bibitem{Domokos94} G.~Domokos, B.~Elliott and S.~Kovesi-Domokos, J.\ Phys.\
{\bf G19} (1993) 899.

\bibitem{Szabo92} A. P. Szabo and R. J. Protheroe, in {\it Proc. Workshop on
High Energy Neutrino Astrophysics,} Honolulu, eds.\ V. J. Stenger \etal~
(World Scientific, Singapore, 1992) p.~24.

\bibitem{Sikora92} M. Sikora and M. Begelman, in {\it Proc. Workshop on High
Energy Neutrino Astrophysics,} Honolulu, eds.\ V. J. Stenger \etal~(World
Scientific, Singapore, 1992) p.~114.

\end{thebibliography}
\end{document}